\documentclass[prd,amssymb,amsmath,amsfonts,nofootinbib,reprint,longbibliography,superscriptaddress, floatfix]{revtex4-2}

\usepackage{graphicx}
\usepackage{lmodern}
\usepackage{amsmath,amssymb}
\usepackage{mathrsfs}
\usepackage{amsfonts}
\usepackage[utf8]{inputenc}
\usepackage{url}
\usepackage[colorlinks]{hyperref}
\usepackage[dvipsnames,x11names,svgnames,rgb,table]{xcolor}
\usepackage{multirow}
\usepackage[normalem]{ulem}
\usepackage{float}
\usepackage{marvosym}
\usepackage{enumerate}
\usepackage{color,soul}
\usepackage{placeins}
\usepackage{bm}
\usepackage{color}
\usepackage{commath}
\allowdisplaybreaks
\usepackage{multirow}
\usepackage{cleveref}
\usepackage{rotating} 
\usepackage{orcidlink}
\usepackage{todonotes}
\usepackage{makecell,tabularx,colortbl}
\usepackage{booktabs,cellspace}
\usepackage{mathtools}

\usepackage{tikz}
\usetikzlibrary{positioning}
\usetikzlibrary{fit}
\usetikzlibrary{shapes.geometric, arrows}

\graphicspath{{plots/}}

\tikzstyle{startstop} = [rectangle, rounded corners,   text centered, draw=black, fill=blue!10]
\tikzstyle{process} = [rectangle, rounded corners, minimum width=2cm, minimum height=1cm, text centered, draw=black, fill=blue!10]
\tikzstyle{decision} = [diamond, aspect=3, minimum width=3cm, minimum height=1cm, text centered, draw=black, fill=red!10]
\tikzstyle{compute} = [rectangle, minimum width=2cm, minimum height=1cm, text centered, draw=black, fill=green!10]
\tikzstyle{estimate} = [rectangle, rounded corners, minimum width=2cm, minimum height=1cm, text centered, draw=black, fill=yellow!10]
\tikzstyle{arrow} = [thick,->,>=stealth]

\newcommand{\hs}{h^{}_s}
\newcommand{\hb}{h^{}_b}

\newcommand{\flow}{f^{}_\mathrm{lower}}
\newcommand{\fup}{f^{}_\mathrm{upper}}

\newcommand{\fecc}{f^{}_\mathrm{ecc}}
\newcommand{\FF}{\mathrm{FF}}
\newcommand{\FFEF}{\mathrm{FF}^{}_{\rm eff}}

\newcommand{\be}{\begin{equation}}
\newcommand{\ee}{\end{equation}}
\newcommand{\bea}{\begin{eqnarray}}
\newcommand{\eea}{\end{eqnarray}}

\newcommand{\qcbanksizeThreeD}{566,974}

\newcommand{\eccbanksizeThreeD}{4,781,475}

\newcommand{\bham}{\affiliation{School of Physics and Astronomy and Institute for Gravitational Wave Astronomy, University of Birmingham, Edgbaston, Birmingham, B15 2TT, United Kingdom}}

\begin{document}

\title{A geometric template bank for the detection of spinning low-mass compact binaries with  moderate orbital eccentricity}

\author{Khun Sang Phukon \orcidlink{0000-0003-1561-0760}}
\email{k.s.phukon@bham.ac.uk}

\author{Patricia Schmidt \orcidlink{0000-0003-1542-1791}}
\email{P.Schmidt@bham.ac.uk} \bham

\author{Geraint Pratten \orcidlink{0000-0003-4984-0775}}
\email{g.pratten@bham.ac.uk} \bham

%~~~~~~~~~~~~~~~ Abstract ~~~~~~~~~~~~~~~
\begin{abstract}
  Compact binaries on eccentric orbits are another class of gravitational-wave (GW) sources that can provide a wealth of information on binary formation pathways and astrophysical environments. However, historically, eccentricity is often neglected in modelled GW searches for compact binaries. We show that currently used modelled searches that employ quasi-circular template banks are highly ineffectual in detecting binary neutron star (BNS) and neutron star--black hole (NSBH) systems with orbital eccentricities in the range of $[10^{-5},0.15]$ at a GW frequency of $15$Hz. For populations of moderately eccentric BNS and NSBH binaries with (anti-)aligned component spins, we demonstrate that quasi-circular template banks fail to detect up to $\sim 40\%$ of such systems. To alleviate these inefficiencies, we develop the first \emph{geometric} template bank for the search of BNSs and NSBH binaries that includes masses, (anti-)aligned spins and moderate eccentricity. Utilising the post-Newtonian inspiral waveform {\tt TaylorF2Ecc} and a global coordinate transformation, we construct a globally flat metric to efficiently place eccentric templates. Our geometric template bank is highly effectual, and significantly improves the recovery of eccentric signals with less than $6\%$ of signals missed due to the finite template spacing in the bank. 
\end{abstract}

%\date{\today}
\maketitle

%%%%%%%%%%%%%%%%%%%%%%%%%%
\section{Introduction}
\label{intro}
%%%%%%%%%%%%%%%%%%%%%%%%%%
Gravitational-wave astronomy has been expanding rapidly through the detection of gravitational waves (GWs) from $\sim 90$ compact binary mergers \cite{LIGOScientific:2021djp,Nitz:2021zwj,Olsen:2022pin,Mehta:2023zlk,Wadekar:2023gea} in the first three observing runs of the Advanced LIGO-Virgo-KAGRA detector network~\cite{LIGOScientific:2014pky, VIRGO:2014yos,Aso:2013eba,KAGRA:2020tym}, which has revealed novel insights into the populations of stellar-mass compact binaries~\cite{KAGRA:2021duu}, the properties of ultra-dense matter~\cite{LIGOScientific:2018cki,LIGOScientific:2019eut} and gravity in the strong-field regime~\cite{LIGOScientific:2021sio}. The recent success of pulsar timing arrays has added another dimension to the progress of GW astronomy and astrophysics by finding the first hints of a nano-hertz stochastic GW background~\cite{NANOGrav:2023gor, EPTA:2023fyk, Reardon:2023gzh, Xu:2023wog}.

The majority of compact binary sources observed by the LIGO-Virgo-KAGRA detectors are found using search pipelines~\cite{Usman:2015kfa, Nitz:2018rgo, Messick:2016aqy, Aubin:2020goo, Chu:2020pjv, Venumadhav:2019tad,Davies:2020tsx,Sachdev:2019vvd,DalCanton:2020vpm,Wadekar:2024zdq} that employ matched filtering~\cite{Sathyaprakash:1991mt, Dhurandhar:1992mw, Owen:1998dk, Allen:2005fk}, which is the optimal technique for detecting well-modelled signals, such as those from coalescing compact binaries, in Gaussian noise. The GW signal from compact binary coalescences can be accurately modelled across a wide range of their parameter space using analytical and numerical techniques (see~\cite{Schmidt:2020ekt} for review on GW modelling techniques). These models are then used to construct a finite grid of waveforms, referred to as the template bank, which the data are filtered against. The limited accuracy of waveform models and the discreteness of the bank lead to a loss of recovered signal-to-noise ratio (SNR). In addition, a lack of physical effects such as general-relativistic spin-precession~\cite{Apostolatos:1994mx, Kidder:1995zr, Schmidt:2010it}, higher-order harmonics~\cite{Pan:2011gk} and orbital eccentricity~\cite{Peters:1963ux} will further reduce the efficacy of the bank and lead to the loss of astrophysically interesting signals. Currently, both of these effects are largely ignored in the template banks used in matched-filter searches.  High-mass signals with these effects are searched for using unmodelled or semi-modelled methods~\cite{Skliris:2020qax,Klimenko:2008fu, Lynch:2015yin}, but these methods are less sensitive in the regime where the total mass $\lesssim 70 M_\odot$~\cite{LIGOScientific:2023lpe}, where modelled searches excel.

Including new degrees of freedom in template banks to account for these effects increases the size of the bank by many orders of magnitude, which makes the search computationally expensive.  New degrees of freedom in the bank offer additional power to recover the GW signal. However, adding new templates to cover the new degrees of freedom enhances the response of the template bank to noise in the detector, which might reduce the sensitivity of search in certain parameter spaces due to the increased false alarm rate~\cite{VanDenBroeck:2009gd,Harry:2016ijz}.  Developing template banks and optimal search methods for eccentric and precessing binaries is challenging and remains an active research area~\cite{McIsaac:2023ijd,Schmidt:2024jbp,Harry:2016ijz, Nitz:2021mzz, Schmidt:2024hac, Dhurkunde:2023qoe, Lenon:2021zac}. 

Despite steady progress in recent years, the analysis frameworks for eccentric binaries are still relatively immature compared to the state-of-the-art for quasi-circular binaries. One reason why the focus has been on non-eccentric binaries is the fact that GW emission very efficiently circularizes the binary's orbit through angular momentum loss~\cite{Peters:1963ux,Peters:1964zz} before they reach the frequencies current ground-based GW detectors are sensitive to. However, $5-10\%$ of binaries formed through dynamical interactions in dense stellar environments may still retain significant orbital eccentricity at a GW frequency of $\sim 20$Hz~\cite{Ye:2019xvf,Samsing:2017xmd,Rodriguez:2018pss}. Detecting eccentric binaries, therefore, would elucidate the formation pathways of binaries~\cite{Zevin:2021rtf} and improve measurement precision by breaking parameter degeneracies~\cite{Yang:2022tig,Sun:2015bva,Xuan:2022qkw,Favata:2013rwa}, with implications for fundamental physics~\cite{Narayan:2023vhm,Saini:2022igm,DuttaRoy:2024aew,Shaikh:2024wyn}. Consequently, there has been a surge in measuring signatures of orbital eccentricity from GW events, and claims that a few events might be eccentric~\cite{Romero-Shaw:2021ual,Romero-Shaw:2022xko,OShea:2021faf,Gupte:2024jfe,Iglesias:2022xfc,Ramos-Buades:2023yhy,Gayathri:2020coq,Wu:2020zwr,Gamba:2021gap,Bonino:2022hkj}. 

Several studies have assessed the effectiveness of quasi-circular template banks in detecting GWs from eccentric binaries, highlighting the need for searches specifically targeting moderately eccentric binaries~\cite{Brown:2009ng, Cokelaer:2009hj, Huerta:2013qb, Wang:2021qsu, Divyajyoti:2023rht, Gadre:2024ndy, Ramos-Buades:2020eju, Lenon:2021zac}. Recent matched-filter searches~\cite{Nitz:2019spj, Nitz:2021vqh, Nitz:2022ltl, Dhurkunde:2023qoe} of LIGO-Virgo-KAGRA data have included eccentricity in template banks using a brute force stochastic method~\cite{Kacanja:2024pjh}; a search for eccentric mergers using particle swarm optimization was also recently developed~\cite{Pal:2023dyg}; un- or semi-modelled methods~\cite{Tiwari:2015gal} were used to search for eccentric binary black hole (BBH) mergers on bound orbits with total binary source mass greater than $ 70 M_\odot$~\cite{LIGOScientific:2023lpe}.  There have also been efforts to search for eccentric BBHs in the lower mass range using the semi-modelled method, where the method is less sensitive~\cite{LIGOScientific:2019dag,KAGRA:2021bhs}. To date, none of these searches have yielded any significant detections of eccentric compact binary mergers. 

In this paper, we present the first \emph{geometric} template bank for matched-filter searches for eccentric low-mass compact binaries with aligned spins by constructing an effective metric based on eccentric the post-Newtonian (PN) inspiral model \texttt{TaylorF2Ecc}~\cite{Moore:2016qxz}. The geometric approach is known to be the optimal strategy for constructing template banks. The resulting eccentric bank is small compared to stochastic eccentric banks in a similar region of parameter space~\cite{Nitz:2019spj,Dhurkunde:2023qoe} , yet highly effective with less than $6\%$ of eccentric signals missed due to the finite grid spacing between templates in the bank. 

The paper is organised as follows. 
In Sec.~\ref{sec:mf}, we give a brief overview of the current status of matched-filter searches and template bank construction methods.  We discuss the waveform model used in this study and the reparametrization of its phase in Sec.~\ref{sec:wf}. We present the effective metric and the flat metric space for lattice placement in Sec.~\ref{sec:metric} and discuss the eccentric template bank construction in Sec.~\ref{sec:bank}. In Sec.~\ref{sec:performance}, we quantify the efficacy of the eccentric bank and demonstrate its superior efficiency in detecting GW signals from spinning low-mass BNS and NSBH with moderate orbital eccentricity. We discuss caveats and directions for future improvements in Sec.~\ref{sec:discussion}. Throughout the paper we work in geometrical units with $G=c=1$.

%%%%%%%%%%%%%%%%%%%%%%%%%%%%%%%%%%%%%%%%%%%
\section{Matched Filtering}
\label{sec:mf}

Current GW modelling techniques can produce the gravitational waveforms emitted by compact binaries to a remarkable accuracy, see e.g.~\cite{Ramos-Buades:2023ehm, Pratten:2020ceb}. These waveforms are characterized by \emph{intrinsic parameters}, e.g. masses, spins, eccentricity, and \emph{extrinsic parameters}, i.e. coalescence phase, coalescence time, distance, sky location and orientation relative to the detector. 
For such modelled signals and under the assumption that the noise is stationary and Gaussian, \emph{matched filtering} is the optimal detection method~\cite{Allen:2005fk} and is widely and successfully used in GW searches~\cite{Usman:2015kfa, Davies:2020tsx, Messick:2016aqy, Adams:2015ulm, Venumadhav:2019tad, LIGOScientific:2019lzm, KAGRA:2023pio}.

The matched filter is defined as the noise-weighted cross-correlation between the observed data stream and the anticipated GW signal~\cite{wiener1949extrapolation,Allen:2005fk}.
Although the matching is conventionally carried out in the frequency-domain, this process produces a timeseries, which is maximized when the template waveform used to filter the data describes a signal present in the data to a very high degree. Normalising the template waveforms appropriately allows to compute the signal-to-noise ratio (SNR).  The peak value of this SNR timeseries is referred to as the SNR associated with the template. 
 
Traditional matched filter searches require a \emph{template bank} that spans (part of) the intrinsic binary parameter space\footnote{Bank-free matched filter-based search methods using particle swarm optimization have been proposed for searches. While they are computationally cheaper than bank-based methods, they provide reduced statistical significance for events~\cite{Wang:2010jma, Normandin:2020ltl, Srivastava:2018wvy, Pal:2023dyg}.}.  Such a bank consists of a set of template waveforms discretely arranged on a finite grid. Each template in the bank is then cross-correlated against the data in each detector, yielding the maximum SNR for the template that best matches the source's parameters.  An SNR exceeding a predetermined threshold indicates the presence of a potential GW signal in data, prompting further signal-consistency tests for confirmation~\cite{Allen:2004gu,Nitz:2017lco,McIsaac:2022odb,Schmidt:2024kxy}. 
Due to the discreteness of the bank, the parameters of the best-matching template may not be precisely the same as those of the source, even if the source lies within the target parameter space covered by the bank. 
This discrepancy induces an intrinsic SNR loss in a discrete template bank. To minimize such a loss in SNR, templates are placed such that at least a specified fraction of the SNR of any possible favourably oriented and located source is recovered. This minimum recoverable fraction of SNR is referred to the {\it minimal match} (MM), which can be related to an event detection loss. 

The MM also serves as an indicator of the finiteness of the template bank, i.e., the ``closeness'' between neighbouring templates in the bank.
Setting a higher MM in the bank increases the number of templates, thereby raising the computational cost of filtering the data. 
Conversely, a lower value reduces computational demands but compromises the detection efficiency of the bank. 
Thus, the construction of a template bank involves a trade-off between computational cost and detection efficiency, making the construction of an optimal bank a challenging task in GW data  analysis~\cite{Manca:2009xw,Allen:2021yuy}.

Several methods for the construction of template banks are available, which can be categorized into four broad approaches: i) geometric~\cite{Babak:2006ty,Cokelaer:2007kx,Brown:2012qf,Harry:2013tca,Roulet:2019hzy}, ii) stochastic~\cite{Babak:2008rb,Harry:2009ea, Ajith:2012mn, Privitera:2013xza, Kacanja:2024pjh},  iii) hybrid~\cite{Fehrmann:2014cpa,Roy:2017oul, Roy:2017qgg, Sharma:2023djw} and iv) random placement~\cite{Messenger:2008ta,Manca:2009xw, Rover:2009wn, Coogan:2022qxs, Schmidt:2023gzj}. 

The \emph{geometric} placement of templates can be viewed as solving the mathematical {\it sphere covering problem} on the waveform manifold~\cite{Prix:2007ks}, where distances between templates or the centers of spheres relate to the mismatch (see Sec.~\ref{sec:geom_approach}).  
The mismatch can be approximated using a second-order Taylor expansion, allowing to define an analytic metric on the waveform manifold~\cite{Owen:1998dk}, which in turn is used to define a regular lattice for placing templates in a locally flat space~\cite{Babak:2006ty,Cokelaer:2007kx,Babak:2012zx}. While effective for inspiral-only waveform models with closed-form expressions~\cite{Dhurandhar:1992mw,Balasubramanian:1995bm}, the geometric method struggles with parameter spaces that have intrinsic curvature, complex boundaries, or high dimensionality~\cite{Manca:2009xw}.

The \emph{stochastic} placement of templates~\cite{Harry:2009ea, Ajith:2012mn} offers a robust alternative to the geometric method. Waveforms are randomly drawn from the parameter space of interest, with new templates added to the bank if they sufficiently differ from existing ones. Unlike the geometric method, this approach does not require a metric over the waveform manifold but instead evaluates the match numerically. Hence, this method can easily be extended to arbitrary parameter spaces. 
However, this method is computationally very intensive, requiring the computation of millions of waveform and matches until no new template is accepted in the bank. 
Importantly, the stochastic method also generates a substantially larger number of templates than the geometric method in a comparable parameter space; it overcovers the space by placing too many redundant templates closely together~\cite{Indik:2017vqq}. 

Considering the parameter space of stellar-mass compact binaries, the geometric method is the optimal way to construct the bank in the low-mass region ($\lesssim 10M_\odot$). 
However, it is less efficient in the high-mass region, where the stochastic method is beneficial. Early attempts to combine these placement methods involved constructing template banks using geometric placement in the low-mass region and stochastic placement in the high-mass region of the parameter space~\cite{Capano:2016dsf,DalCanton:2017ala}.
\emph{Hybrid} approaches combine the efficiency of geometric placement with the robustness of the stochastic method across the entire parameter space~\cite{Fehrmann:2014cpa,Roy:2017qgg,Roy:2017oul,Sharma:2023djw}. 

The \emph{random}  placement of templates is an emerging template placement strategy for compact binary searches originally developed for searches of continuous gravitational waves~\cite{Messenger:2008ta}. Unlike other methods, this placement method does not require exhaustive computations of matches between templates to add a new template to the bank, and all templates in this method are randomly sampled from a probability distribution given by the metric of waveform manifold until convergence is achieved or stopping criteria are reached~\cite{Coogan:2022qxs, Schmidt:2023gzj}. Random placement generates more templates than stochastic ones, but the bank construction can be orders of magnitude faster.

Here, we develop the first geometric template bank for spinning, eccentric binaries. Dedicated matched-filter searches for eccentric BNS, NSBH and sub-solar mass binaries have so far only used template banks generated using stochastic methods~\cite{Nitz:2019spj, Dhurkunde:2023qoe, Nitz:2022ltl,Nitz:2021mzz}. Eccentric stochastic template banks contain about two orders of magnitude more templates than a quasi-circular, nonspinning bank that covers the same mass parameter space would~\cite{Nitz:2019spj,Dhurkunde:2023qoe}.
We will show in Sec.~\ref{sec:performance} that our geometric template bank for eccentric, spinning binaries only leads to an increase in the number of templates by one order of magnitude while remaining highly efficient.

%%%%%%%%%%%%%%%%%%%%%%%%%%%%%%
\section{Waveform Model}
\label{sec:wf}
%%%%%%%%%%%%%%%%%%%%%%%%%%%%%%
Several waveform models for GW signals from compact binaries on eccentric orbits exist in the literature and they can broadly be grouped as follows: 1) inspiral-only waveforms~\cite{Moore:2016qxz,Moore:2019xkm,Tiwari:2019jtz,Tanay:2016zog,Huerta:2014eca,Klein:2018ybm,Klein:2021jtd,Arredondo:2024nsl} developed within the PN approximation to general relativity\footnote{The PN formalism approximates the general relativistic description of a system by systematically adding corrections to the Newtonian dynamics as a polynomial in orbital velocity $v^2$. An $n$PN order description of the system includes a term proportional to $v^{2n}$ relative to the leading term in the relevant expressions.}~\cite{Blanchet:2013haa}; 
2) surrogate models derived from numerical relativity (NR) simulations of eccentric binaries, providing accurate waveform predictions beyond the PN approximation~\cite{Islam:2021mha}; 
3) effective-one-body (EOB)-based~\cite{Buonanno:1998gg,Buonanno:2000ef} inspiral-merger-ringdown (IMR) waveforms~\cite{Cao:2017ndf,Liu:2021pkr,Chiaramello:2020ehz,Nagar:2021gss,Khalil:2021txt,Ramos-Buades:2021adz,Liu:2023ldr}; 4) IMR waveforms models constructed using various phenomenological approaches~\cite{Paul:2024ujx,East:2012xq,Huerta:2016rwp,Huerta:2017kez,Hinder:2017sxy,Setyawati:2021gom,Wang:2023ueg} and 5) gravitational self-force (GSF) models for eccentric extreme mass ratio inspirals~\cite{Lynch:2021ogr,Katz:2021yft}. In this work, we use the inspiral-only PN waveform model {\tt TaylorF2Ecc}~\cite{Moore:2016qxz}, which is an analytic frequency domain model, obtained via the stationary phase approximation (SPA)~\cite{Cutler:1994ys}. The model is available via {\tt LALSimulation} routines of the LIGO Algorithm Library~\cite{lalsuite}.

The {\tt TaylorF2Ecc} waveform model extends the quasi-circular {\tt TaylorF2} model~\cite{Buonanno:2009zt,Arun:2008kb} to include leading-order eccentricity effects for binaries with comparable masses and orbital eccentricities up to $e [10 {\rm Hz}] \lesssim  0.2$~\cite{Favata:2021vhw}. It builds on the quasi-Keplerian (QK) description~\cite{Damour:1988mr,SCHAFER1993196} of the conservative dynamics of PN elliptical orbits and equations that account for dissipation in orbital elements due to GW emission. 

The QK formalism provides parameterized solutions to the conservative orbital dynamics of eccentric binaries in terms of three eccentricity variables, $e^{}_t$ (associated with the orbital period), $e^{}_r$ (associated with the radial motion), $e^{}_\phi$ (associated with the azimuthal motion), in combination with various other orbital elements. All of these eccentricity variables $(e^{}_t, e^{}_r, e^{}_\phi)$ are related to each other through conserved orbital energy and angular momentum ({\it see} Eq.~(345) of~\cite{Blanchet:2013haa}) but depend on the choice of coordinate system\footnote{There are proposals for gauge-independent definitions of eccentricity using GW or orbital frequencies at apocenter and pericenter passages, inspired by earlier works~\cite{Ramos-Buades:2022lgf,Bonino:2022hkj,Shaikh:2023ypz,Mora:2002gf,Mora:2003wt,Bonino:2024xrv,Boschini:2024scu}. These definitions help infer eccentricity from GW observations and NR simulations~\cite{Ramos-Buades:2019uvh,Bonino:2022hkj,Ramos-Buades:2023yhy,Bonino:2024xrv}. For additional definitions, see Ref.~\cite{Will:2019lfe} and references therein.}. 
For convenience, the temporal eccentricity variable $e^{}_t$ is defined as the eccentricity parameter $(e\equiv e^{}_t)$ in the {\tt TaylorF2Ecc} waveforms, which reduces to the Newtonian definition of eccentricity $e^{}_{\mathrm{Newt}}$ in the appropriate limit and allows to reproduce quasi-circular expressions when $e^{}_t \rightarrow 0$.

The {\tt TaylorF2Ecc}  waveform leverages the 3PN QK description of the conservative orbital dynamics~\cite{Memmesheimer:2004cv,Konigsdorffer:2006zt} for the computation of the orbit averaged energy and angular momentum fluxes, and the evolution of orbital elements under 3PN gravitational radiation reaction for non-spinning binaries~\cite{Arun:2009mc}.  With these inputs,  the leading order secular, eccentric phase corrections, at $O(e^2)$, are included at each PN order up to the 3PN  term in the 3.5 PN accurate quasi-circular phase of GW radiation.  Contributions to the phase from rapid, oscillatory variations in the orbital elements due to the eccentricity of the orbit are ignored as they minimally affect the waveform in the LIGO-like detectors' band.  The quasi-circular part of the phase includes spin corrections to 3.5PN order but does not account for spin-eccentricity cross terms. 
The amplitude of {\tt TaylorF2Ecc} is  Newtonian order accurate with no eccentricity correction in it.
Additionally, {\tt TaylorF2Ecc}  uses a 3PN accurate analytical expression to model the decay of eccentricity with GW frequency $[e^{}_t(f)]$, which is accurate in the small eccentricity limit (see Eq.~(4.17) of \cite{Moore:2016qxz}). 
Efforts are ongoing to accurately model the eccentricity evolution as a function of frequency~\cite{Bonino:2024xrv} and to include spin-eccentricity coupling effects in the GW phase~\cite{Henry:2023tka,Klein:2021jtd, Klein:2018ybm,Klein:2010ti,Sridhar:2024zms}.  
The omission of the oscillatory modulation in the phase and the use of a 3PN accurate analytical expression for the evolution of eccentricity restrict the validity of the {\tt TaylorF2Ecc} waveform model to small eccentricities.  
Also, as pointed out in Ref.~\cite{Moore:2016qxz}, QK solutions start diverging from GSF calculations as binaries become more asymmetric; hence, this waveform model is more reliable for comparable mass binaries.

A waveform from the {\tt TaylorF2}-family, derived through the SPA, is given in the detector frame as
\be
\label{eq:waveform}
\tilde{h}(f;  \vec{\rm \theta}) = \mathcal{A}(f;\vec{\rm \theta}) f^{-7/6}e^{i\left( \Psi_{\mathrm{F}2}(f;\vec{\rm \theta}) - \pi/4 \right) },
\ee
where $ \vec{\rm \theta}$ represents the set of binary parameters, which are broadly categorized into intrinsic $ \vec{\rm \theta}_{\rm int}$ and extrinsic $ \vec{\rm \theta}_{\rm ext}$  parameters. The intrinsic parameters $ \vec{\rm \theta}_{\rm int}$ include masses $m^{}_{(1,2)}$, the $z$-component of each spin $s^{}_{(1,2)z}$ and the orbital eccentricity $e^{}_0$ at a fiducial frequency $\fecc$ : $ \vec{\rm \theta}_{\rm int}=\{ m^{}_1, m^{}_2, s^{}_{1z}, s^{}_{2z}, e^{}_0  \} \subset  \vec{\rm \theta}$. The extrinsic parameters $\vec{\rm \theta}_{\rm ext}$ are the luminosity distance $D^{}_L$, inclination angle $\iota$, right ascension $\alpha$, declination $\delta$, polarization angle $\psi$, arrival time at the detector $t^{}_0$ and the corresponding phase $\phi^{}_0$: $\vec{\rm \theta}_{\rm ext} = \{ D^{}_L, \iota, \alpha, \delta, \psi, t^{}_0, \phi^{}_0 \} \subset \vec{\rm \theta}$. The {\tt TaylorF2Ecc}   phase $\Psi^{}_{\mathrm{F}2e}$ can be expressed as
\be
\label{eq:phase1}
\Psi^{}_{\mathrm{F}2e}(f; \vec{\rm \theta}) =  2\pi f t^{}_0  - 2\phi^{}_0 +  \Psi^{}_{\mathrm {QC}} (f; \vec{\rm \theta}_{\rm int})+ \Psi^{}_{\mathrm {Ecc}}(f; \vec{\rm \theta}_{\rm int}) ,
\ee
where $\Psi^{}_{\mathrm {QC}} (f; \vec{\rm \theta}_{\rm int})$ is the quasi-circular phase and $\Psi^{}_{\mathrm {Ecc}} (f; \vec{\rm \theta}_{\rm int})$ represents the eccentricity correction. The amplitude $\mathcal{A}$ of {\tt TaylorF2Ecc} is truncated at the Newtonian order~\cite{Cutler:1994ys} with no eccentric corrections and is given as
\be
\mathcal{A}(f;\vec{\rm \theta}) = \sqrt{\frac{5}{24}} \frac{\mathcal{M}^{-5/3}}{\pi^{2/3}D^{}_L},
\ee
 where $\mathcal{M} = (m^{}_1 m^{}_2)^{3/5}/(m^{}_1 + m^{}_2)^{1/5}$ is the chirp mass.

The quasi-circular phase $\Psi^{}_{\mathrm {QC}}$ is accurate up to 3.5PN order with spin corrections  and can be expanded in powers of the GW frequency $f$ in the following form~\cite{Arun:2008kb,Buonanno:2009zt} 
\be
\label{Eq:qc_phase1}
\begin{split}
\Psi^{}_{\mathrm {QC}} (f; \vec{\rm \theta}_{\rm int}) =&~\sum_{i=0}^{7}  \varphi^{}_i (\vec{\rm \theta}_{\rm int}) f^{(-5+i)/3} + \\
&\sum_{i=5}^{6}   \varphi^{\ell}_i (\vec{\rm \theta}_{\rm int})\log{(f)}\, f^{(-5+i)/3},
\end{split}
\ee
where $i$ is twice the PN order.
Each of the nine PN coefficients $\{ \varphi^{}_0,\, \varphi^{}_2,\, \varphi^{}_3,\, \varphi^{}_4,\, \varphi^{}_5,\, \varphi^{}_6,\, \varphi^{}_7,\, \varphi^{\ell}_5,\, \varphi^{\ell}_6 \}$ in Eq.~\eqref{Eq:qc_phase1} has a distinct frequency dependency and is a function of the component masses and spins of the binary. The complete expressions for these PN coefficients are given in Appendix~\ref{app:pn_coeff}.   

Similarly, we can   express  the  3PN accurate eccentric phase $\Psi_{\rm Ecc}$ (Eq.~(6.26) in Ref.~\cite{Moore:2016qxz}) as follows
\be
\label{Eq:ecc_phase1}
\begin{split}
\Psi^{}_{\mathrm {Ecc}} (f; \vec{\rm \theta}_{\rm int})  =&~ \sum_{i,j=0}^{6}  \varepsilon^{}_{ij} (\vec{\rm \theta}_{\rm int}) f_{\rm ecc}^{(19+3j)/9} f^{(-34+3i)/9}  + \\
& \varepsilon^\ell_{60} (\vec{\rm \theta}_{\rm int}) \log{(f)}\, f_{\rm ecc}^{19/9}f^{-16/9}  +\\
& \varepsilon_{06}^\ell (\vec{\rm \theta}_{\rm int})  \log{(\fecc)} \,  f_{\rm ecc}^{37/9}f^{-34/9} ,
\end{split}
\ee
where $\fecc$ is the GW frequency at which the initial eccentricity $[e^{}_0 \equiv e^{}_t(\fecc)]$ of the binary system is specified, and the sum of the indices $\left(i+j\right)$ is twice the PN order.  Unlike the quasi-circular PN coefficients, the coefficients in Eq.~\eqref{Eq:ecc_phase1}  are independent of the spin parameters. 

The eccentric phase has 19 PN coefficients: one Newtonian  coefficient  $\varepsilon^{}_{00}$, two  coefficients at 1PN $\{ \varepsilon^{}_{20},  \varepsilon^{}_{02} \}$, two  coefficients at 1.5PN $\{\varepsilon^{}_{30},  \varepsilon^{}_{03}\}$,  three  coefficients at 2PN $\{ \varepsilon^{}_{40}, \varepsilon^{}_{22},  \varepsilon^{}_{04}\}$,  four coefficients at 2.5PN $\{ \varepsilon^{}_{50},   \varepsilon^{}_{23},   \varepsilon^{}_{32},   \varepsilon^{}_{05} \}$,  seven coefficients 3PN  $\{ \varepsilon^{}_{60},   \varepsilon^{}_{24},   \varepsilon^{}_{33},   \varepsilon^{}_{42},   \varepsilon^{}_{06},  \varepsilon^{\ell}_{60},   \varepsilon^{\ell}_{06} \}$; the complete expressions are given in Appendix~\ref{app:pn_coeff_ecc}.

Following Ref.~\cite{Brown:2012qf}, we express Eq.~\eqref{Eq:qc_phase1} together with the phase- and time-shift factors from Eq.~\eqref{eq:phase1} as a series in the dimensionless frequency parameter $x\equiv f/f^{}_0$, where $f_0^{}$ is set at a  fiducial value of $70 {\rm Hz}$. This choice mitigates any potential numerical overflow and renders the coefficients of  $x$ dimensionless. The resulting expression for the quasi-circular phase can be written as
\be
\label{eq:qc_phase_x}
\begin{split}
\Psi_{\mathrm {QC}} ^\prime(x; \vec{\rm \theta}) ~&:= 2\pi ft^{}_0 - 2 \phi^{}_0 + \Psi^{}_{\mathrm{QC}} \\
&=-2\phi^{}_0 +\sum_{i=0}^{8}  \zeta^{}_i (\vec{\rm \theta}) x^{(-5+i)/3} +  \\
&\quad \quad\sum_{i=5}^{6}   \zeta^{\ell}_i (\vec{\rm \theta}_{\rm int})\log{(x)}\, x^{(-5+i)/3},
\end{split}
\ee
where the eight dimensionless coefficients are
\be
\begin{split}
&\zeta^{}_0 = \varphi^{}_0 f_0^{-5/3},\,  \zeta^{}_2 = \varphi^{}_2/ f^{}_0,\,  \zeta^{}_3 = \varphi^{}_3 f_0^{-2/3},\\
&\zeta^{}_4 = \varphi^{}_4 f_0^{-1/3},\,   \zeta^{}_6 = (\varphi^{}_6 + \varphi^{\ell}_6 \log f_0^{}  )f_0^{1/3},\, \zeta^{}_7 = \varphi^{}_7 f_0^{2/3} \\
&\zeta^\ell_5 =  \varphi_5^\ell,\, \zeta^\ell_6 =  \varphi^\ell_6   f_0^{1/3},\,   \zeta^{}_8 = 2\pi t^{}_0 f^{}_0.
\end{split}
\ee

In Eq.~\eqref{eq:qc_phase_x}, we show the coefficients in the power series of $x^{-5/3}$ up to order 8.  We absorb the  $\varphi^{}_5$  coefficient and any other terms that have no dependencies in variables $x$ or $f$ into the overall phase shift factor $\phi^{}_0$ following~\cite{Pai:2012mv, Brown:2012qf,Ohme:2013nsa, Tanaka:2000xy}. The overall  time-shift $t^{}_0$ parameter is absorbed in the coefficients $\zeta^{}_8$.  

Similarly, Eq.~\eqref{Eq:ecc_phase1} can be recast in terms of $x$ as follows
\be
\label{eq:ec_phase_x}
\begin{split}
\Psi'_{\mathrm {Ecc}} (x; \vec{\rm \theta}_{\rm int}, \fecc)  =&~ \sum_{i=0}^{6}  \kappa^{}_{i} (\vec{\rm \theta}_{\rm int}, \fecc)  x^{(-34+3i)/9}  + \\
& \kappa^\ell_{6} (\vec{\rm \theta}_{\rm int}, \fecc) \log{(x)}\, x^{-16/9}.
\end{split}
\ee
Here  $\Psi'_{\mathrm {Ecc}}$ is  equivalent to  $\Psi^{}_{\mathrm {Ecc}}$ but differs in the power series variable and   the $\kappa$-coefficients are combinations of the PN coefficients  $\varepsilon^{}_{ij}$s from Eq.~\eqref{Eq:ecc_phase1}.    The general form of $\kappa^{}_i $ for $\Psi^{}_{\mathrm {Ecc}}$, accurate up to  $n$PN order, is  
\be
\kappa^{}_i  \approx  \sum^{2n-j}_{j=0}\varepsilon^{}_{ij}  (\vec{\rm \theta}_{\rm int}) f_{\rm ecc}^{(19+3j)/9} f_0^{(-34+3i)/9}.
\ee
The coefficient $\kappa^\ell_{6}$ arises when the eccentric phase 
is given to 3PN order and expressed as 
\be
 \kappa_6^\ell =   \varepsilon_{60}^\ell    f_{\rm ecc}^{19/9}  f_0^{-16/9}. 
\ee

The total phase of the GW waveform $\Psi^{}_{\mathrm{F}2e}$ can be reparametrized with a total of 16 coefficients $ \Xi \coloneqq \{ \zeta^{}_i,\, \zeta_i^\ell,\,  \kappa^{}_i,\,  \kappa^\ell_{6} \}$ along with  the constant phase shift $\phi^{}_0$ as 
\be
\label{eq:dimensionless_phase}
\Psi^{}_{\mathrm{F}2e}(x;\phi^{}_0, \Xi ) =  \Psi_{\mathrm {QC}} ^\prime(x; \phi^{}_0, \zeta^{}_i , \zeta_i^\ell) + \Psi'_{\mathrm {Ecc}} (x;\kappa^{}_i,   \kappa^\ell_{6} ).
\ee
We treat all $\Xi$-parameters except $\zeta^{}_8$ as a new set of intrinsic parameters, i.e.,
\begin{align}
\label{eq:intr_coord_N}
    \vec{\rm \theta}'_{\rm int} =& \{\zeta^{}_0,\, \zeta^{}_2,\, \zeta^{}_3,\, \zeta^{}_4,\, \zeta^{}_6,\, \zeta^{}_7,\,  \zeta_5^\ell,\,   \zeta_6^\ell,\, \\ \nonumber
    & \kappa^{}_0,\,   \kappa^{}_2,\,  \kappa^{}_3,\,  \kappa^{}_4,\, \kappa^{}_5,\,  \kappa^{}_6,\, \kappa^\ell_{6}  \}. 
\end{align}

The remaining parameters, $\phi^{}_0$ and $\zeta^{}_8$ are the new  extrinsic parameters $\vec{\rm \theta}'_{ext} = \{\phi^{}_0,\, \zeta^{}_8 \}$. 
We note that the specific choice of these parameters depends on the PN order being used for the waveform.

In the rest of the paper, we will use this new parametrization for $\Psi^{}_{\mathrm{F}2e}$  with 3.5PN accurate  $\Psi'_{\mathrm {QC}}$  and  $\Psi'_{\mathrm {Ecc}}$ at 2PN order, {\it i.e.}, 
\begin{equation}
    \Psi_{\rm{F}2e}(x; \vec{\theta}'_{\rm int}, \vec{\theta}'_{\rm ext}) \equiv \Psi'^{3.5\rm PN}_{\rm QC} +  \Psi'^{2\rm PN}_{\rm Ecc}.
\end{equation}
Because of the termination of $ \Psi'_{\rm Ecc}$ at 2PN order,  the coefficients $\{ \kappa^{}_5,\,  \kappa^{}_6,\, \kappa^\ell_{6} \}$ are omitted making   the $\Xi$-parameter space   $13$-dimensional.  We also dropped   contributions of PN terms higher than 2PN in $\{\kappa^{}_0,\,   \kappa^{}_2,\,  \kappa^{}_3,\,  \kappa^{}_4\}$ to make the coeffiecients consistent with the 2PN consideration of the eccentric phase. For completeness, we give expressions of all dimensionless $ \{\kappa^{}_i,  \kappa^\ell_{6}\}$ coefficients of Eq.~\eqref{eq:dimensionless_phase}, containing terms to 3PN order in Appendix~\ref{app:dimensionless_phase}.

%%%%%%%%%%%%%%%%%%%%%%%%%%%%%%%%%%%%%%%%%%%%%%%%%%%%%%%%
\section{A metric for Eccentric Signals}
\label{sec:metric}
%%%%%%%%%%%%%%%%%%%%%%%%%%%%%%%%%%%%%%%%%%%%%%%%%%%%%%%%

%%%%%%%%%%%%%%%%%%%%%%%%%%%%%%%%%%%%%%%%%%%%%%%%%%%%%%%%
\subsection{Geometric Approach}
\label{sec:geom_approach}
%%%%%%%%%%%%%%%%%%%%%%%%%%%%%%%%%%%%%%%%%%%%%%%%%%%%%%%%

We present and employ a \emph{geometric} method to construct a bank of eccentric waveforms for matched-filtering spanning the intrinsic parameter space $(m^{}_1, m^{}_2, s^{}_{1z}, s^{}_{2z}, e^{}_0)$, building on the works of~\cite{Owen:1995tm, Owen:1998dk,Babak:2006ty,Brown:2012qf, Harry:2013tca,Tanaka:2000xy}. 

Naively, a template bank can be viewed as a grid of discrete points in parameter space. However, it is non-trivial to determine the optimal grid. The geometric approach defines a metric on the waveform space, which, in combination with a local flatness theorem, provides a prescription for choosing grid points. The placement depends on two crucial factors:

\begin{enumerate}
\item The curvature of the waveform manifold: 
It impacts how the spacing between waveforms is measured, thus affecting the distribution of templates. A flat manifold ensures adequate coverage and minimal redundancy. 
\item The coordinate choice: Coordinates can be curvilinear even if the manifold itself is flat or nearly flat. A ``Cartesian-like'' (or close to) coordinate frame is preferred for effective template placement. This is akin to laying a regular grid on a plane using either Cartesian or polar coordinates, where the choice of coordinates can simplify or complicate the lattice regularity.
\end{enumerate}
These geometric properties of a waveform manifold are described by its metric, which depends on the waveform model and its parametrization. Matched-filtering naturally induces a metric for the waveform manifold. Generally, the similarity between two waveforms $h^{}_1$ and $h^{}_2$ is quantified in terms of the \emph{overlap} $\mathcal{O}$, defined as
\begin{equation}
\label{Eq:overlap}
 \mathcal{O}(h^{}_1,h^{}_2) =  (h^{}_1|h^{}_2) \coloneqq    4 \, \mathrm{Re} \int^{\fup}_{\flow } df \dfrac{\tilde{h}^{}_1(f)\tilde{h}_2^*(f)}{S^{}_n(f)},
\end{equation}
where $\tilde{h}^{}_1$ and $\tilde{h}^{}_2$ are the Fourier transforms of $h^{}_1$ and $h^{}_2$, $\flow (\fup)$ is the lower(upper)-cutoff frequency of a detector's sensitive range and $S^{}_n(f)$ denotes the one-sided power spectral density of detector noise.

Waveform differences reduce the overlap. Therefore, it is a measure of the recoverable fraction of the optimal SNR. In template bank construction, it is customary to normalize waveforms such that  $(h|h) = 1$. In this paper, we will consider normalized waveforms unless otherwise specified. This implies that two identical waveforms have $\mathcal{O}=1$, whereas it is zero if two waveforms are orthogonal to each other.

If the same waveform model describes two waveforms $h^{}_1=h(\vec{\rm \theta})$ and $h^{}_2=h(\vec{\rm \theta}+\Delta \vec{\rm \theta})$, then the overlap is sensitive to the relative parameter differences $\Delta \vec{\theta}$. For small $\Delta \vec{\rm \theta}$, the overlap  can then be approximated as a second-order Taylor expansion of $\Delta \vec{\rm \theta}$ 
\begin{align}
    \mathcal{O} (h_1, h_2) &=  (h(\vec{\rm \theta}) \mid h(\vec{\rm \theta} + \Delta \vec{\rm \theta})) \\ \nonumber
    &\approx 1 + \frac{1}{2} \left(  \frac{\partial^2 \mathcal{O}}{\partial \theta^\alpha \partial \theta^\beta} \right) \Delta \theta^\alpha  \Delta \theta^\beta. 
\end{align}
The terms on right hand side of the above equation define a metric $g_{\alpha \beta}$~\cite{Owen:1995tm, Owen:1998dk} 
\be
\label{eq:metric_full}
 g^{}_{\alpha \beta} (\vec{\rm \theta}) :=  -\frac{1}{2} \left(  \frac{\partial^2 \mathcal{O}}{\partial \theta^\alpha \partial \theta^\beta} \right),
\ee
which is equivalent to the Fisher information matrix (FIM).

In matched-filter searches or for quantifying the closeness of two waveforms in the intrinsic parameter space, it is more common to use the {\it match} $\mathrm{M}$, which is the overlap maximized over the extrinsic parameters $\vec{\rm \theta}_{ext} = (t^{}_0, \phi^{}_0)$~\cite{Sathyaprakash:1991mt}
\begin{align}
\label{eq:match_no_metric}
\mathrm{M}(h^{}_1, h^{}_2) =  \underset{\vec{\rm \theta}_{ext}}{\max}\,\mathcal{O} (h^{}_1, h^{}_2). 
\end{align}
For two waveforms that are located nearby in the intrinsic parameter space, the match can be approximated via the metric of the intrinsic parameter space, $\Gamma^{}_{ij}$, which can  be obtained by projecting the full metric $g^{}_{\alpha \beta}$ (Eq.~\eqref{eq:metric_full}) onto the subspace of intrinsic parameters:
\begin{equation}
\label{eq:projection}
\Gamma^{}_{ij} = g^{}_{i i} - g^{a b} g^{}_{i a} g^{}_{b j},
\end{equation}
where the indices $\{a, b\} $ correspond to  the extrinsic parameters  $ \vec{\theta}_{\rm ext}$ and indices $\{i,j\}$  are for intrinsic parameters $\vec{\rm \theta}_{\rm int}$. 
The form of the metric $\Gamma^{}_{ij} $ is the Schur complement of sub-matrix $g^{}_{ab}$ related to the extrinsic parameters within the larger symmetric matrix $g^{}_{\alpha \beta}$. The inverse of the sub-matrix $g^{}_{\alpha \beta}$ is denoted by $g^{\alpha \beta}$. 
Using the intrinsic metric $\Gamma^{}_{ij}$, the {\it match} between waveforms $h_1$ and $h_2$ can be approximated as
\begin{equation}
\label{eq:match}
\mathrm{M}(h_1 ,h_2) \approx  1 - \Gamma^{}_{ij} \Delta \theta^i_{\rm int} \Delta \theta^j_{\rm int}.
\end{equation}

The stochastic method for constructing a template bank is iterative and consists of evaluating the match between waveforms in the bank and trial waveforms. If their match is below a desired match threshold, referred to as the \emph{minimal match} $\mathrm{MM}$, the trial waveform will be added to the bank. Typically, this requires millions of match computations. The approximation of the match in terms of the intrinsic metric as given in Eq.~\eqref{eq:match} offers a quick way to evaluate the match if $\Gamma^{}_{ij}$ can be determined quickly.

%%%%%%%%%%%%%%%%%%%%%%%%%%%
\subsection{Flat metric space}
\label{sec:coord}
%%%%%%%%%%%%%%%%%%%%%%%%%%%
In the canonical intrinsic waveform parameter space, the elements of $\Gamma^{}_{ij}$ vary rapidly across the parameter space due to the complex dynamics of binary systems, leading to significant intrinsic curvature. Large curvature in the metric space implies that the evaluation of Eq.~\eqref{eq:match} depends on both positions and the separation of templates. 
Templates need to be placed irregularly and densely to satisfy the MM criterion.  Therefore, we look for a better coordinate system (or parameter space), where the intrinsic metric $\Gamma^{}_{ij}$ does not vary significantly or, ideally, remains constant across the parameter space.

To alleviate the effects of large curvature in the metric space, the chirp time coordinate space has been used to construct banks for non-spinning binaries as these variables are almost Cartesian~\cite{Sathyaprakash:1991mt,Dhurandhar:1992mw}. 
In particular, the Newtonian ($\tau^{}_0$) and the 1.5PN ($\tau^{}_3$) contributions to the chirp time are used to parametrize PN waveforms yielding a locally flat metric~\cite{Owen:1998dk, Babak:2006ty,Cokelaer:2007kx}. 
In general, however, it is difficult to identify a waveform parametrization that leads to a locally or globally flat metric. In Ref.~\cite{Ajith:2012mn}, the dimensionless chirp time coordinates were generalised to include non-precessing spins.
Recently, Ref.~\cite{Roulet:2019hzy} proposed a new method that decomposes waveforms on an orthonormal basis and uses the basis coefficients of the phase to construct an Euclidean metric. 

In Refs.~\cite{Brown:2012qf, Harry:2013tca, Ohme:2013nsa}, it was shown that the dimensionless coefficients of the PN expansion parameter $x$ in the quasi-circular {\tt TaylorF2} phase, i.e. ($\zeta^{}_i, \zeta_i^\ell$ in Eq.~\eqref{eq:qc_phase_x}), can constitute a globally flat metric for non-precessing binaries on quasi-circular orbits.  
However, this metric space is eight-dimensional compared to the four-dimensional physical coordinate space. Large dimensionality and curvilinearity make template placement difficult in the phase coefficient parameter space. The issues arising from this are dealt with by transforming into a Cartesian coordinate system and then applying a dimensionality reduction algorithm to find a lower dimensional space (see Sec.~\ref{sec:placement}).

Here, we construct a globally flat metric for the eccentric, aligned-spin waveform model {\tt TaylorF2Ecc} (see Sec.~\ref{sec:wf}) via the PN phase coefficients $ \Xi := \left( \zeta^{}_i, \zeta_i^\ell, \kappa^{}_i \right) $ as given in Eqs.~\eqref{eq:qc_phase_x} and~\eqref{eq:ec_phase_x}, and follow the prescription introduced in~\cite{Owen:1995tm, Owen:1998dk} for obtaining the metric $\Gamma^{}_{ij}(\Xi)$ (Eq.~\eqref{eq:projection}). 

To do this,  we first derive a metric, $g^\prime_{\alpha \beta}$, to approximate the overlap $\mathcal{O}$  that is  maximized  over   $\phi^{}_0$: $\mathcal{O}'=\underset{\phi^{}_0}{\max}\,\mathcal{O}$. This metric $g^\prime_{\alpha \beta}$ represents the 13 dimensional $\Xi$-space, which is  orthogonal to $\phi_0^{}$ of  the 14 dimensional $\left(\Xi, \phi_0^{}\right)$-space.   The coordinates of the 13 dimensional $\Xi$-space denoted by Greek indices  $\Xi_\alpha^{}$ includes 12  the intrinsic parameters $\vec{\theta'}$ from Eq.~\eqref{eq:intr_coord_N} and one extrinsic parameter $\zeta^{}_8$.  The $\phi^{}_0$-maximized overlap $\mathcal{O}'$ in the  $\Xi$-space  is  related to the proper distance between two points and given by
\be
1-\mathcal{O}' = g^\prime_{\alpha \beta} \Delta\Xi^\alpha \Delta \Xi^\beta,
\ee
where $\Delta\Xi^\alpha$ are differences in the parameters of two waveforms.  Following~\cite{Owen:1998dk}, we derive the metric $g^\prime_{\alpha \beta}$ using  the noise moment functionals $\mathcal{J}$ for the gradients $\psi^{}_\alpha = \frac{\partial \Psi^{}_{\mathrm{F2e}} }{\partial \Xi^\alpha}$ of the waveform phase $\Psi^{}_{\mathrm{F2e}}$ with respect to the $\Xi^{}_\alpha$ coordinates. Using the moment functionals the metric $g'_{\alpha \beta}$ is expressed as 
\be
\label{eq:metric_temporal_component}
g^\prime_{\alpha \beta} \left(\Xi_\alpha\right) = \frac{1}{2} \left(  \mathcal{J} [\psi^{}_\alpha \psi^{}_\beta ] - \mathcal{J}[\psi^{}_\alpha] \mathcal{J}[\psi^{}_\beta] \right).
\ee
The moment functionals $\mathcal{J}$ can be expanded in terms of the  moments of the detector's PSD~\cite{Poisson:1995ef},
which we define similarly to~\cite{Keppel:2013kia} but not quite in the same way as other papers~\cite{ Poisson:1995ef, Owen:1995tm} as 
\begin{align}
\label{eq:moments}
I(q,l) &:= \displaystyle\int_{x^{}_{L}}^{x^{}_{U}} \frac{x^{-q / 3} \log^{l}(x)}{S^{}_n(f^{}_0 x)} dx, \\
\label{eq:moments_ratio}
J(q,l) &:= \frac{I(q,l)}{I(7,0)}.
\end{align}
 The integrals $I(q,l)$ represent  moments of the PSD, where $q$ and $l$ are powers to $x^{-1/3}$ and $\log(x)$, respectively.  The quantities $x^{}_{L} = \flow/f^{}_0$  and $x^{}_{U} =  \fup/f^{}_0$ specify the minimum and the maximum frequency of the detector's sensitive band. Using moments of PSD, moment functionals for an  input function $a(x)$ can be  given as
\be
\label{eq:normalizedmoments}
\mathcal{J}[a(x)] := \frac{1}{I(7,0)} \int_{x^{}_{L}}^{x^{}_{U}} \frac{x^{-7/3}}{S^{}_n(f^{}_0 x)}
a(x) dx.
\ee

 In deriving metric elements $g'_{\alpha \beta}$, the derivatives of the waveform phase $\psi^{}_\alpha $  with respect to $\Xi^{}_\alpha$,  and their products $\psi^{}_\alpha \psi^{}_\beta$ are passed as input functions $a(x)$ to the moment functionals.  These  input functions $a(x)$ have the general form
 \be
 \label{eq:general_expression}
 a(x) = x^{q/3}\log^l(x).
 \ee
 The logarithmic factor in Eq.~\eqref{eq:general_expression} arises when $a(x)$ are derived for phase coefficients of waveforms with logarithmic terms.  The simple form of the input function $a(x)$ allows evaluation corresponding moment functional  $\mathcal{J}[a(x)]$ by using the ratio of moment functions $J$, as given in ~\eqref{eq:moments_ratio} 
\be
\label{eq:normalizedmomentsexpanded}
\mathcal{J} \left[   a(x) \right]  = \mathcal{J} \left[   x^{q/3} \log^l(x) \right] =   J(7-q,l).
\ee

As previously noted, we input  $\psi^{}_\alpha$ and  $\psi^{}_\alpha \psi^{}_\beta$   to Eq.~\eqref{eq:metric_temporal_component} and Eq.~\eqref{eq:normalizedmomentsexpanded} to derive the metric $g'_{\alpha \beta}$.  We provide expressions for $\psi^{}_\alpha$  in Table~\ref{tab:phase_coeff_derivatives}, but we do not  display  $\psi^{}_\alpha \psi^{}_\beta$ terms  as they are too numerous. These input functions derived for given $\Xi^{}_\alpha$ do not depend on any physical parameters or on  $\Xi_\alpha$ values. If $x^{}_{L},~x^{}_{U}$  are held fixed throughout the parameter space, the  metric $g'_{\alpha \beta}$ is invariant in the $\Xi$-space, resulting in a flat metric space.  We fix  $x^{}_{L},~x^{}_{U}$   at   $\flow=15 {\rm Hz}$, $\fup = 1000 {\rm Hz}$ and and $f^{}_0 = 70 {\rm Hz}$ to get rid of  any  curvature-related complexities in  $\Xi$-space.

\begin{table}[t]
\centering
\begin{ruledtabular}
\begin{tabular}{lccr}
 $\quad\Xi_\alpha $ & $\psi^{}_\alpha = \frac{\partial \Psi_{\mathrm{F2e}}}{\partial \Xi^\alpha}$ & $q$ & $l\quad$ \\
\hline
$\quad\zeta^{}_0$ & $x^{-5/3}$ &  -5 &  $0\quad$\\
$\quad\zeta^{}_2$ & $x^{-1}$ & -3 &  $0\quad$\\
 $\quad\zeta^{}_3$ & $x^{-2/3}$ &-2 & $0\quad$\\
 $\quad\zeta^{}_4$ & $x^{-1/3}$ & -1&$0 \quad$\\
  $\quad\zeta^{}_6$& $x^{1/3}$ &  1& $0\quad$\\
     $\quad\zeta^{}_7$& $x^{2/3}$ & 2& $0\quad$\\
 $\quad\zeta^{}_8$&  $x$  & 3 &  $0\quad$\\
          $\quad\zeta_5^\ell$& $\log x$ & 0& $1\quad$\\
     $\quad\zeta_6^\ell$& $x^{1/3} \log x$ & 1& $1\quad$\\
     $\quad\kappa^{}_0$ & $x^{-34/9}$ & -34/3 & $0\quad$ \\
          $\quad\kappa^{}_2$ & $x^{-28/9}$ & -28/3 & $0\quad$ \\
               $\quad\kappa^{}_3$ & $x^{-25/9}$ & -25/3 & $0\quad$ \\
                    $\quad\kappa^{}_4$ & $x^{-22/9}$ & -22/3 & $0\quad$ \\
%\hline
\end{tabular}
\end{ruledtabular}
 \caption{  \label{tab:phase_coeff_derivatives}   Table for derivatives $\psi^{}_\alpha$  of waveform phase $\Psi^{}_{\mathrm{F}2e}(\Psi^{\prime 3.5\mathrm{PN}}_{\mathrm {QC}},\Psi_{\mathrm {Ecc}}^{\prime2\mathrm{PN}})$  with respect to  13  $ \Xi^{}_\alpha$ coordinates  given in the first column.  The second column of the table shows the derivatives  $\psi^{}_\alpha$, and each derivative takes the form of $a(x)$ given in Eq.~\eqref{eq:general_expression}. The corresponding exponents $q$ and $l$ in $a(x)$ for a $\psi^{}_\alpha$ are shown in the third and fourth columns, respectively. For brevity, we do not give the expressions of cross terms of two derivatives $\psi^{}_\alpha \psi^{}_\beta$, which also takes the form of $a(x)$  given in this table.}
\end{table}

After flattening the $\Xi$-space, we derive  the  metric $\Gamma^{}_{ij}$ for the 12 dimensional (12D)  subspace of  intrinsic parameters $\vec{\theta}'_{\rm int} := \{ \Xi_i\} =  \{\zeta^{}_0, \zeta^{}_2, \zeta^{}_3, \zeta^{}_4, \zeta^{}_6, \zeta^{}_7,  \zeta_5^\ell, \zeta_6^\ell, \kappa^{}_0, \kappa^{}_2, \kappa^{}_3, \kappa^{}_4 \}$ by projecting out the temporal component $\Xi_0 = \zeta_8 = 2 \pi f^{}_0 t^{}_0$ from the metric space given by $g^\prime_{\alpha \beta}$. The resulting effective metric in the  12D intrinsic parameter space is 
\be
    \label{eq:maximised_metric}
\Gamma^{}_{ij}\left(\Xi_i\right)= g^\prime_{ij} - g^{\prime00}g^\prime_{i0}g^\prime_{j0},
\ee
where the zero index corresponds to the temporal coordinate $\zeta^{}_8$ and Latin indices denote the intrinsic parameters $\{ \Xi_i^{}\}$. This effective, globally constant  metric  $\Gamma^{}_{ij}$ approximates the match between two waveforms in the 12D $\Xi_i^{}$-space. % as

%%%%%%%%%%%%%%%%%%%%%%%%%%%%%%%%%
\subsection{Efficient, Lower-Dimensional   Metric Space}
\label{sec:placement}
%%%%%%%%%%%%%%%%%%%%%%%%%%%%%%%%% 
For geometric template  placement, merely finding a coordinate system that flattens the parameter space is insufficient. For constructing an effectual and optimal geometric template bank, it is essential that the metric-based match (Eq.~\eqref{eq:match}) in the flat metric space approximates the numerical match with high accuracy. This ensures efficient template placement and adequate coverage of the parameter space. The metric approximation of the match is highly accurate for quasi-circular aligned-spin waveforms in a flat metric space. However, it fails to reliably predict the match between waveforms when higher-order PN terms such as tidal terms or non-GR effects are considered to construct a flat space~\cite{Kalaghatgi:2015nia,Harry:2021hls,Sharma:2023djw}.  
Keeping this in mind, we conduct a test to ensure that this not the case when including the eccentric terms and that the metric match in the 12D $\Xi^{}_i$-space is indeed an excellent approximation of the numerical match before proceeding with the template bank construction. 

\begin{figure}
\includegraphics[width=.5\textwidth]{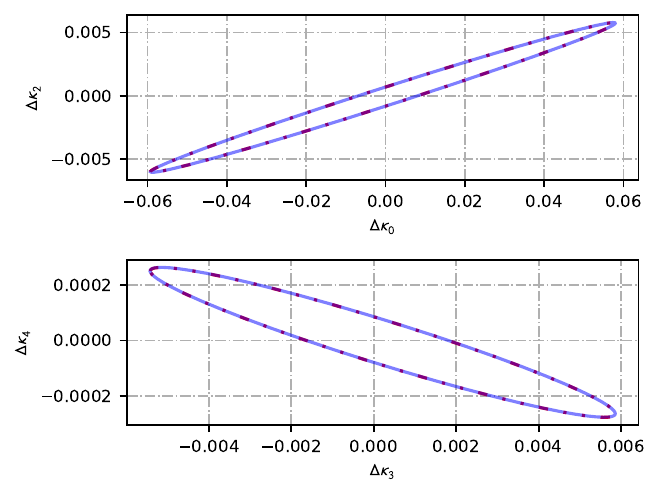}
\caption{\label{fig:match_compare_kappa_bns} Ambiguity ellipses on differences in  $\kappa_i$  $\left(  \Delta \kappa_i =  \kappa_i ^{\rm waveform1}- \kappa_i ^{\rm waveform2}\right)$ of {\tt TaylorF2Ecc} waveforms with match $0.935$ in the BNS parameter space. The red dots are obtained from the numerical evaluation of the match between waveforms, and the blue, solid ellipses are computed using the metric approximation of the match. The metric-based and numerical match computations agree perfectly in $\kappa_i$-space.  We also find perfect agreement in other combinations of $\kappa_i$.}
\end{figure}

For this test, we select random {\tt TaylorF2Ecc} waveform pairs with component masses $\leq~3 M^{}_\odot$ and a numerical match of ${\rm M}^{}_{\rm num} \simeq 0.935$. This match value is close to the match between two templates in a template bank with $\mathrm{MM} = 0.97$. For these waveforms with coordinate differences $\Delta \Xi^{}_i$, those match values are recovered with errors $\sim 10^{-5}$ using the metric approximation of match,  $\left(1 - \Gamma^{}_{ij} \Delta \Xi^{i} \Delta \Xi^{j}\right)$ in the $\Xi_i^{}$-space.  We show the {\it ambiguity ellipses}\footnote{We call thess ellipses as {\it ambiguity ellipses} following terminologies of~\cite{Owen:1995tm}. These ellipses are projections of the $(n-1)$ dimensional ellipsoids obtained from Eq.~\eqref{eq:match} in an $n$ dimensional parameter space. A waveform on these ellipsoidal surface with parameters  $\theta^{i}_{\rm int} + \Delta \theta^{i}_{\rm int} $ has a match ${\rm M}$ with a reference waveform at $\theta^{i}_{\rm int}$. } in Fig.~\ref{fig:match_compare_kappa_bns} for $(\kappa_0^{}, \kappa_2^{})$ and  $(\kappa_3^{}, \kappa_4^{})$ plane, where each coordinate axis denotes differences between corresponding $\Xi_i$-coordinates of the waveforms. We observe similar agreement for ellipses in other $\Xi_i$ coordinates as well.

It is clear that the metric approximation of match accurately predict the numerical match between waveforms in the $\Xi_i$-space. A key revelation from Fig.~\ref{fig:match_compare_kappa_bns}  is that the $\Xi^{}_i$ coordinates are not orthogonal.  The shape and orientation of the ellipses are determined by the metric in the flat $\Xi_i^{}$ space. The ellipses in Fig.~\ref{fig:match_compare_kappa_bns} are not aligned along the coordinate axes, indicating the non-orthogonality of coordinates. It is easier to place templates or lattice points in a Cartesian or orthonormal coordinate system where lattice or grid regularity can be maintained throughout the parameter space, and the distance between points becomes the Euclidean distance. 
Since the metric does not depend on the parameters $ \Xi^{}_i =  \{  \zeta^{}_i, \zeta_i^\ell, \kappa^{}_i\}$, it is possible to perform a coordinate transformation to an orthonormal coordinate system. We first transform to a Cartesian coordinate and then do a scaling such that the metric becomes the identity matrix in the new coordinates.  
The orthonormal coordinates are given by~\cite{Brown:2012qf,Harry:2013tca}
\be
\mu^{}_i = V_{ij} \sqrt{E^{i}_i}\, \Xi^{j},
\ee
where $V^{}_{ij}$ denotes the $j^{\rm th}$ component of the $i^{\rm th}$ eigenvector and $E^{i}_{i}$ the eigenvalues corresponding to $i^{\rm th}$ eigen vector  of the matrix $\Gamma^{}_{ij}$. 

The new coordinates $\mu^{}_i$  constitute a 12D Cartesian space where the metric is the identity matrix, but it is sub-optimal to place lattice points in the flat space of such dimensionality~\cite{Prix:2007ks}. To deal with the increased dimensionality of the space, we apply a principal component analysis (PCA) to find an effective lower-dimensional space of the Cartesian coordinates $\mu^{}_i$ for the placement of the lattice points to construct the template similarly to~\cite{Brown:2012qf,Harry:2013tca,Ohme:2013nsa, Sharma:2023djw}. 
To determine the principal components $\xi^{}_i$,  we first compute the $\mu^{}_i$ coordinates from one million uniformly distributed random binary parameters covering ranges given in Table~\ref{tab:params} and their covariance matrix. The eigenvectors  $C^{}_{ij}$ of the covariance matrix form an orthogonal basis and indicate principal directions along which the parameter space vary.
We  project the coordinates $\mu^{}_i$ onto the orthogonal basis given by $C^{}_{ij}$ to obtain the principal components $\xi^{}_i$ as 
\be
\xi^{}_i = C^{}_{ij}\mu^{j}.
\ee

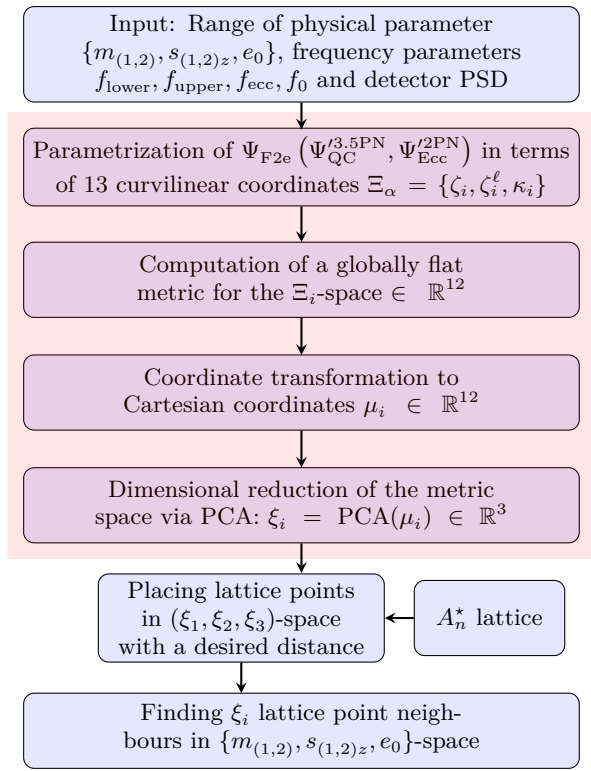
\begin{figure}\centering 
	\begin{tikzpicture}[node distance=1.5cm]
	
	\node (in) [startstop, text width=.4\textwidth] {Input: Range of physical parameter  $\{ m^{}_{(1,2)},  s^{}_{(1,2)z},  e^{}_0\}$,   frequency parameters $\flow, \fup, f_{\rm ecc}, f_0^{}$ and detector PSD \par };
	
	\node (wf) [process, text width=.4\textwidth, below of=in] {Parametrization of $\Psi^{}_{\mathrm{F2e}}\left(\Psi'^{3.5\mathrm{PN}}_{\mathrm {QC}},\Psi_{\mathrm {Ecc}}'^{2\mathrm{PN}}\right)$ in terms of 13 curvilinear coordinates $\Xi^{} _\alpha = \{ \zeta^{}_i , \zeta_i^\ell,  \kappa^{}_i  \}$   \par };
	
	\node(metric)[process, text width=.4\textwidth, below of=wf]{Computation of a globally flat metric for the $\Xi^{}_i$-space $\in \mathbb{R}^{12}$\par};
	
	\node(cartesiancoord)[process, text width=.4\textwidth, below of=metric]{Coordinate transformation to Cartesian coordinates $\mu^{}_i \in \mathbb{R}^{12}$\par};
	
	\node(pcaspace)[process, text width=.4\textwidth, below of=cartesiancoord]{Dimensional reduction of the metric space via PCA:  $\xi_i^{}={\rm PCA}(\mu^{}_i)\in \mathbb{R}^{3}$};
		
	\node(latticepoints)[process,  below of=pcaspace, text width=.2\textwidth, xshift=-.8cm]{Placing lattice points in  $(\xi^{}_1,\xi^{}_2,\xi^{}_3)$-space with a desired distance};

	\node(lattice)[process, right of=latticepoints, xshift=1.8cm]{$A_n^\star$ lattice};
	
	\node(physicalpoints)[process, text width=.4\textwidth, below of=latticepoints, xshift=.8cm]{Finding $\xi_i^{}$ lattice point neighbours in  $\{ m^{}_{(1,2)},  s^{}_{(1,2)z},  e^{}_0\}$-space};

	\node(box1)[draw, red!10, solid,fit=(wf)(metric)(cartesiancoord) (pcaspace), inner sep=2mm , fill=red, fill opacity=0.1] {};
	\draw[arrow] (in) -- ++(0,-1) (wf);
	\draw[arrow] (wf) --++(0,-1) (metric);
	\draw[arrow]  (metric)--++(0,-1)(cartesiancoord);
	\draw[arrow]  (cartesiancoord)--++(0,-1)(pcaspace);
	\draw[arrow]  (pcaspace)--++(0,-.9)(latticepoints);
	\draw[arrow] (lattice)--(latticepoints);
	\draw[arrow] (latticepoints)--++(0,-1)(physicalpoints);;
	
	\end{tikzpicture}
	
	\caption{Block diagram showing the key steps for generating an eccentric template bank with the  geometric approach. The blocks in the red-filled region indicate the steps involved in computing the global, flat metric for a lower dimensional space obtained through PCA. The blue blocks at the bottom show the steps for placing the templates.}
	\label{fig:bankgen}
\end{figure}

The  $\xi^{}_i$ coordinates provide a 12D Euclidean space system, where mismatches between waveforms $h_1^{}$, $h_2^{}$ are squared distances between corresponding points in the principal component space, given as: 
\begin{equation}
\label{eq:match_xi}
1- \mathrm{M}(h^{}_1, h^{}_2) = \sum^{n \leq 12}_{i=1}  \Delta \xi^2_i.
\end{equation}
While one could in principle use all 12 $\xi_i^{}$ coordinates to compute mismatches between waveforms, this is inefficient, as the first few components typically capture  most of  variations of waveform manifold in the the physical range of masses, spins and eccentricity considered here.  Among these $\xi_i^{}$ coordinates, $\xi_1^{}$ is the most significant  coordinate that captures the  majority of variance in mismatch between waveforms in the physical parameter space and $\xi_{12}^{}$ is the least significant coordinate.  Using first few significant $\xi^{}_i$ coordinates, the numerical mismatch between waveforms  can be accurately approximated  using Eq.~\eqref{eq:match_xi} within an acceptable  error tolerance. These reduced set of  significant coordinates form a lower-dimensional,   effective metric space.  We show the flow of constructing the lower-dimensional effective metric space for {\tt TaylorF2Ecc} waveform in red-filled region of  Fig.~\ref{fig:bankgen}. In next section, we will show that a three dimensional effective metric space in $\left( \xi_1^{}, \xi_2^{}, \xi_3^{}\right)$ can accurately represent the {\tt TaylorF2Ecc} waveforms in the physical parameter space considered here.

%%%%%%%%%%%%%%%%%%%%%%%%%%%%%%%%%%%%%%%%%%%%%%%%%%%%%%%%%%%%%%%%%%%%%%%%%%%%
\section{Geometric template bank for the {\tt TaylorF2Ecc} metric}
\label{sec:bank}
%%%%%%%%%%%%%%%%%%%%%%%%%%%%%%%%%%%%%%%%%%%%%%%%%%%%%%%%%%%%%%%%%%%%%%%%%%%%

\begin{table}[t]
    \centering
    \begin{ruledtabular}
    \begin{tabular}{c c}
        Parameter & Range \\
        \hline 
        Primary   mass $(m^{}_1)$ &  $[1, 7] M_\odot$ \\
        Secondary  mass $(m_2^{})$  & $[1,3] M_\odot$ \\
        BH spin magnitude   &  $[0.0, 0.9]$\\
        NS spin magnitude   &  $[0.0, 0.05]$ \\
        Orbital eccentricity  $\left(e_0^{}=e(15{\rm Hz}) \right)$ & $[10^{-5}, 0.15]$\\
        Lower-cutoff frequency  $(\flow)$ & $15\,$Hz \\
        Upper-cutoff frequency  $(\fup)$ & $1000\,$Hz \\
        PSD & {\tt aligoO4\_high.txt}~\cite{noise_curves}\\
    \end{tabular}
    \end{ruledtabular}
    \caption{Summary of the physical parameter space of eccentric binaries considered here for the geometric, {\tt TaylorF2Ecc} bank construction.  The primary components of the binaries include of both BHs and NSs, with any object of mass $\leq 3 M_\odot$ classified as NS. The secondary component parameters are restricted to NSs. Additionally, the {\tt aligoO4\_high.txt} is the file  of the projected optimistic noise curve of the LIGO detectors for the O4 observation run.  The lower-cutoff frequency  $(\flow)$ and upper-cutoff frequency $(\fup)$ are other key inputs for constructing the template bank.}
    \label{tab:params}
\end{table}

\begin{figure}
\includegraphics[width=0.5\textwidth]{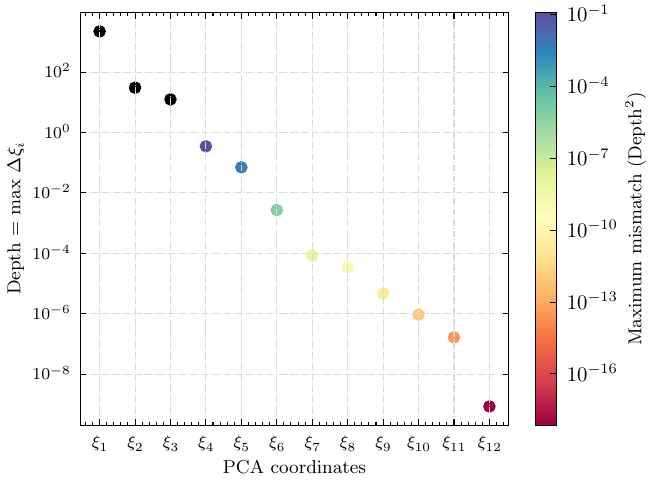}
\caption{\label{fig:depth} Depth or width of our targeted physical parameter space along each $\xi^{}_i$ direction. The colours correspond to the the maximum mismatch considering the variability of the parameter space along a single $\xi^{}_i$ direction except for the first three $\xi^{}_i$ directions. We do not compute the maximum mismatches for $\left(  \xi^{}_1, \xi^{}_2, \xi^{}_3\right)$ as the depths for these directions are significantly larger, making the metric-based approximation of the match (Eq.~\eqref{eq:match_xi}) inappropriate for binaries at opposite ends of any of these three directions.} 
\end{figure}

We construct a template bank that covers a nominal parameter space spanned by BNS and low-mass NSBH binaries without considering tidal effects. The parameter ranges are given Table~\ref{tab:params}.  
To determine the number of required principal components for our targeted physical parameter space, we draw $10^6$ random samples and compute the corresponding $\xi$-values following the method described in Sec.~\ref{sec:placement}. We then calculate the depth along each $\xi^{}_i$-direction, defined as the difference between the maximum and minimum values of each $\xi^{}_i$ and expressed as   
\be
{\rm Depth} \coloneqq \mathrm{max}\,\Delta\xi_i^{}=max(\xi^{}_i) - min(\xi^{}_i).
\ee
The depth along each PCA direction is shown in Fig.~\ref{fig:depth}. As each $\xi$ is a combination of mass, spin and eccentricity parameters, the depth along each $\xi$ reflects the variability of the parameter combinations across the parameter space. 
We find substantial depth across the first three PCA directions, $\left( \xi^{}_1, \xi^{}_2, \xi^{}_3 \right)$ compared to the remaining ones, indicating that the first three PCA components are the most critical to cover the parameter space sufficiently.

We further assess the importance of the remaining coordinates $\xi^{}_4$ to $\xi^{}_{12}$ for covering the parameter space by computing the match between binaries in the $\xi$-space using equation Eq.~\eqref{eq:match_xi}. 
To do so, we assume that parameter combinations between two binaries are such that they are at opposite ends of the specific $\xi^{}_i$-coordinate under investigation, while the other $\xi$ coordinates are identical. The resulting mismatch between such binaries is the square of depth along the particular $\xi_i$-direction, given by $\mathrm{M} = 1-{\rm Depth}^2$. It corresponds to the minimum attainable match, $\mathrm{M}_\mathrm{min}$,  accounting for variation only along that direction. 
The colourbar in Fig.~\ref{fig:depth} shows the corresponding maximum mismatch value for each $\xi$-direction, $\xi^{}_4-\xi^{}_{12}$, given by $1-\mathrm{M}_\mathrm{min}$.  
The maximum mismatch values for binaries at extreme ends of those $\xi_i^{}$ directions are less than $0.1$. If we were to place lattice points along the $\xi_4^{}-\xi_{12}^{}$ directions, these would need to be gridded. However, given the small mismatches between extreme points along those directions, the grid spacing would be unresponsive to variations in physical parameters and to changes in the match between templates. We conclude that $\xi_4^{}-\xi_{12}^{}$ do not contribute enough to enhance the accuracy, but they significantly increase the dimensionality of the metric space used for lattice placement, therefore we exclude them from the template bank construction.  

Our targeted parameter space has significant depths along the first three directions, $\xi^{}_1, \xi^{}_2, \xi^{}_3$, which is why the match between two extreme points along these directions is not computed using the geometric method, as the Fisher-based computation fails for such large separations. Consequently, we argue that the first three directions provide a sufficient representation of the signal variation over our targeted parameter space. This representation is akin to other low-rank approximation methods used to reduce the matched-filtering cost~\cite{Cannon:2010qh,Field:2011mf,Kulkarni:2018mkc}.

We now examine how accurately the three most significant principal directions approximate the match over the parameter space.  To do this, we compare the numerical match, denoted $\mathrm{M}_{\rm num}$ and given by Eq.~\eqref{eq:match_no_metric}, with the metric-based match $\mathrm{M}_{\rm apx}$  given in Eq.~\eqref{eq:match_xi} for a large number of binaries within our target parameter space  given in Table~\ref{tab:params}. The binary parameters covering the parameter ranges are distributed uniformly. We choose each binary pair such that $\mathrm{M}_{\rm num} \geq 0.98$, and compute the metric-based match at different levels of principal component truncation. 
 
\begin{figure}
\includegraphics[width=0.5\textwidth]{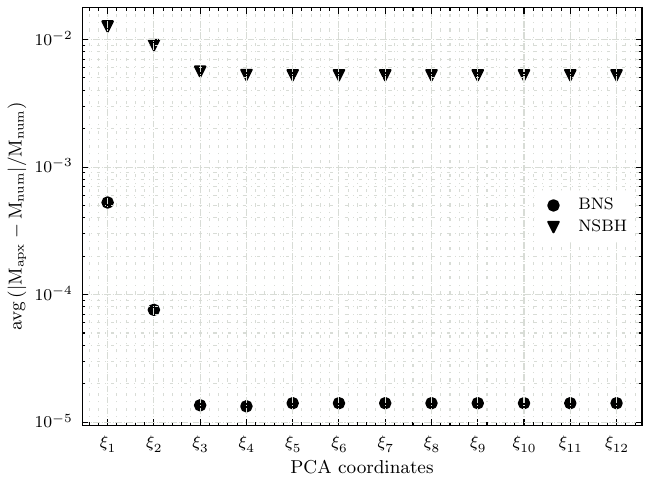}
\caption{\label{fig:cumulative_match} Average error in the metric-based match $\mathrm{M}_{\rm apx}$  relative to the numerical match $\mathrm{M}_{\rm num}$ for two populations of binaries as a function of the number of principal components retained in Eq.~\eqref{eq:match_xi}. For both uniformly distributed BNS and NSBH populations covering the target parameter space given in Table~\ref{tab:params}, the binary pairs for the match computation are chosen such that $\mathrm{M}_{\rm num} \geq 0.98$.  The relative error in $\mathrm{M}_{\rm apx}$ reaches a saturation level after the inclusion of the first three principal components, indicating that the remaining principal components are negligible. }
\end{figure}

We expect two main sources of error in $\mathrm{M}_{\rm apx}$. The first one is the truncation error due to selecting a subset of principal components. In Fig.~\ref{fig:cumulative_match}, we show the average percentage error in the match due to principal component truncation for NSBH (triangles) and BNS (circles) subpopulations.  We see that the error decreases rapidly as the number of principal components increases for the first three components $\xi^{}_1,\, \xi^{}_2,\, \xi^{}_3$, reaching saturation thereafter.  This indicates that any waveform in the target parameter space can be represented accurately by combinations of the first three PCA components, $(\xi^{}_1,\, \xi^{}_2,\, \xi^{}_3)$.  

\begin{figure}
\includegraphics[width=0.5\textwidth]{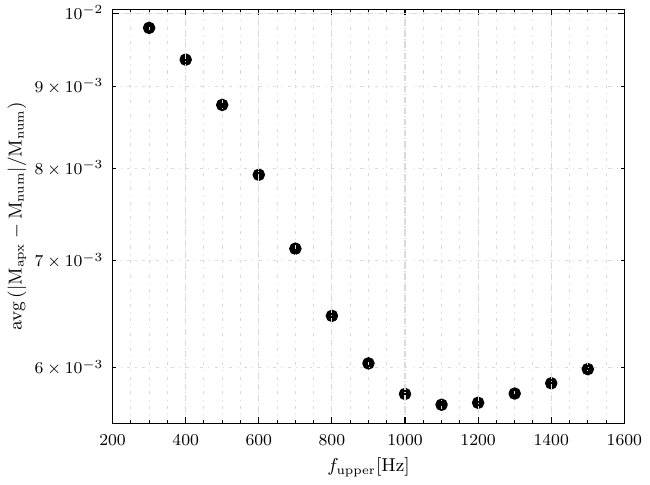}
\caption{\label{fig:match_fupper} Figure demonstrates  the dependence of accuracy of the metric-approximated  match $\mathrm{M}_{\rm apx}$  on the choice of the upper cut-off frequency $f^{}_{\rm upper}$ in the metric  evaluation. All principal components are utilized. % for computing  $\mathrm{M}_{\rm apx}$ for metric varying across $f^{}_{\rm upper}$. 
The average errors in $\mathrm{M}_{\rm apx}$ relative to the numerical match $\mathrm{M}_{\rm num}$ are calculated for the combined population used in Fig.~\ref{fig:cumulative_match}.  }
\end{figure}

The second source of error in the match calculation arises from over- or underestimating the bandwidth of the signals when deriving the metric. In our derivation of the globally flat metric, we fixed the upper cutoff frequency to $\fup = 1000\, {\rm Hz}$. For BNS systems, mergers typically occur at frequencies above $1000\, {\rm Hz}$. However, this bandwidth truncation has a negligible effect for low-mass binaries as the sensitivity of current-generation detectors decreases rapidly at high frequencies. 
In contrast, NSBH systems can merge below $1000$ Hz, leading to larger disagreements of $\mathrm{M}_{\rm apx}$ with $\mathrm{M}_{\rm num}$ for NSBHs compared to BNS systems as can also be seen in Fig.~\ref{fig:cumulative_match}. The impact of bandwidth truncation in computing the flat metric on the accuracy of $\mathrm{M}_{\rm apx}$   is illustrated in  Fig.~\ref{fig:match_fupper} by varying $\fup$ between 300-1500 Hz. The average errors in $\mathrm{M}_{\rm apx}$, taking into account all principal components in its evaluation for varying $\fup$, are computed for the combined population used in Fig.~\ref{fig:cumulative_match}.

For GW detection, we want the fractional average loss in the match to be less than the desired minimal match of our template bank, which we choose to be $\rm MM = 0.97$. With at least three principal components and $\fup = 1000 {\rm Hz}$, the fractional error is orders of magnitude smaller than the desired MM for both subpopulations. This validates our choices of principal components and upper cutoff frequency, making them appropriate to use in the bank construction. While the fiducial 
 $\fup=1000$Hz, used in our metric is not the value ($\fup=1100$Hz) that minimizes the errors in Fig~\ref{fig:match_fupper}, the resulting differences are negligible. Thus, our results would remain consistent even if we had chosen the value of $\fup=1100$ Hz.  It is also worth noting that a cutoff frequency $\fup$ larger than the actual termination frequency of the waveform makes the $\mathrm{M}_{\rm apx}$ overly sensitive to variations in the physical parameters and causes an over-coverage of templates in the NSBH region~\cite{Harry:2013tca}.

After identifying the most significant principal components or $\xi^{}_i$ coordinates, we place a $A^*_n$ lattice~\cite{anStarBook,Prix:2007ks, Wette:2014tca} in the 3D $\xi_1^{}-\xi^{}_2-\xi^{}_3$ space to cover our target parameter space. In three dimensions, the $A^*_n$ lattice provides a truncated octahedron lattice system. Lattice points are iteratively filled until no more can be added, i.e., until the spacing between all points corresponds to a desired minimal match of $0.97$.  In principle, one can generate higher dimensional lattices using the remaining higher order $\xi^{}_i$, but the gain would be negligible at the cost of putting many redundant lattice points close together or over-covering the lattice space. 

Although we can easily place the lattice points in the principal component coordinate space, $\xi_1^{}-\xi_2^{}-\xi_3^{}$, there is no analytical inverse mapping to obtain the corresponding coordinates in the physical parameter space. Therefore, we follow the brute force method carried out in previous studies~\cite{Brown:2012qf}. For a given lattice point, this method generates random points in the physical parameter space and calculates their distance with the lattice point in $\xi$-space. A random point in the physical parameter space is considered to be a solution when the distance is less than a pre-defined tolerance, $\epsilon$. For the bank presented in this work, we choose $\epsilon = 10^{-2}$, corresponding to a mismatch of $1 - {\rm M}_{\rm apx} \sim10^{-4}$.  
This process is computationally challenging and intensive.  To parallelise the process, we split the physical parameter space into non-overlapping chirp mass bins and use the binary search algorithm KD Tree~\cite{Virtanen:2019joe} to find the nearest random point. The final steps involved in the bank construction are summarised in the bottom blue boxes of Fig.~\ref{fig:bankgen}. 

\begin{figure*}[h]
\centering
\includegraphics[width=0.9\textwidth, height=.4\textheight ]{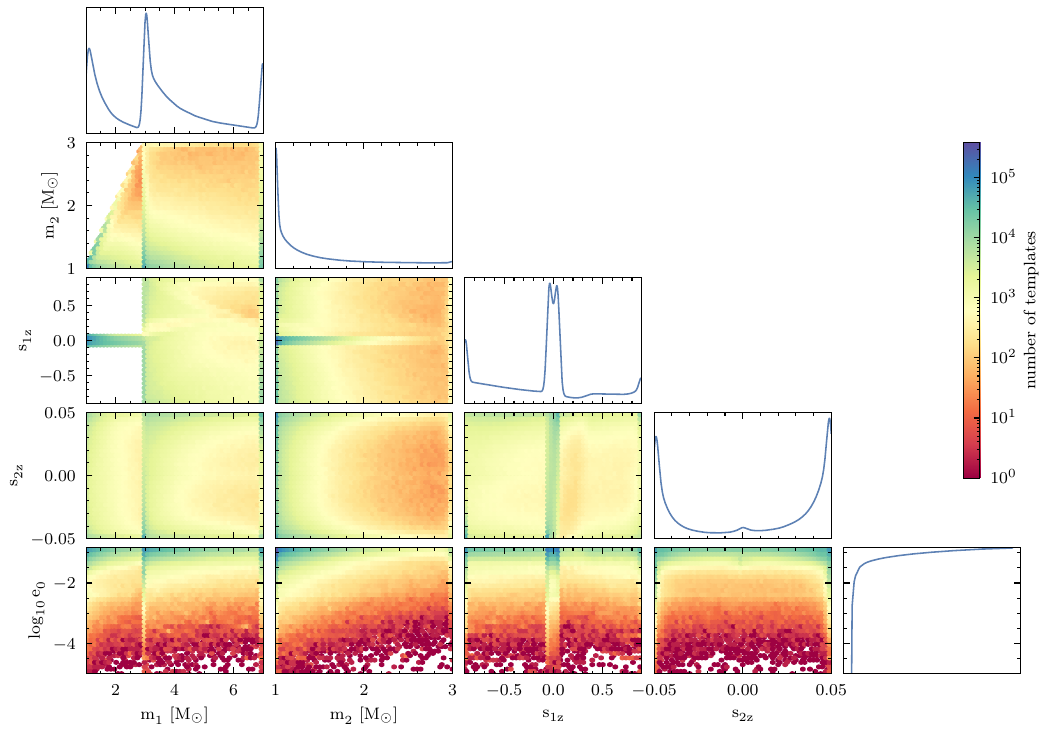}
\caption{\label{fig:ecc_bank_component_mass} Template density in hexagonal binning of different pairs of template parameters of the eccentric template bank. The {\tt TaylorF2Ecc}-based geometric eccentric bank covers the parameter space $m_1 \in [1,7]\, \mathrm{M}_\odot$, $m_2 \in [1,3] \,\mathrm{M}_\odot$,  $s^{}_{1z} \in [-0.9,0.9]$, $s^{}_{2z} \in [-0.05, 0.05]$, $e_0 \in [10^{-5}-0.15]$, where eccentricity is defined at a GW frequency of 15Hz.
The MM of the bank is $0.97$ and the templates are placed using the three-dimensional $A^*_n$ lattice~\cite{Prix:2007ks}. The blue lines show the marginal distributions of the template density for each bank parameter. The colors represent the number of templates in each hexagon.}
\end{figure*}

\begin{figure*}[h]
\centering
\includegraphics[width=0.9\textwidth, height=.4\textheight ]{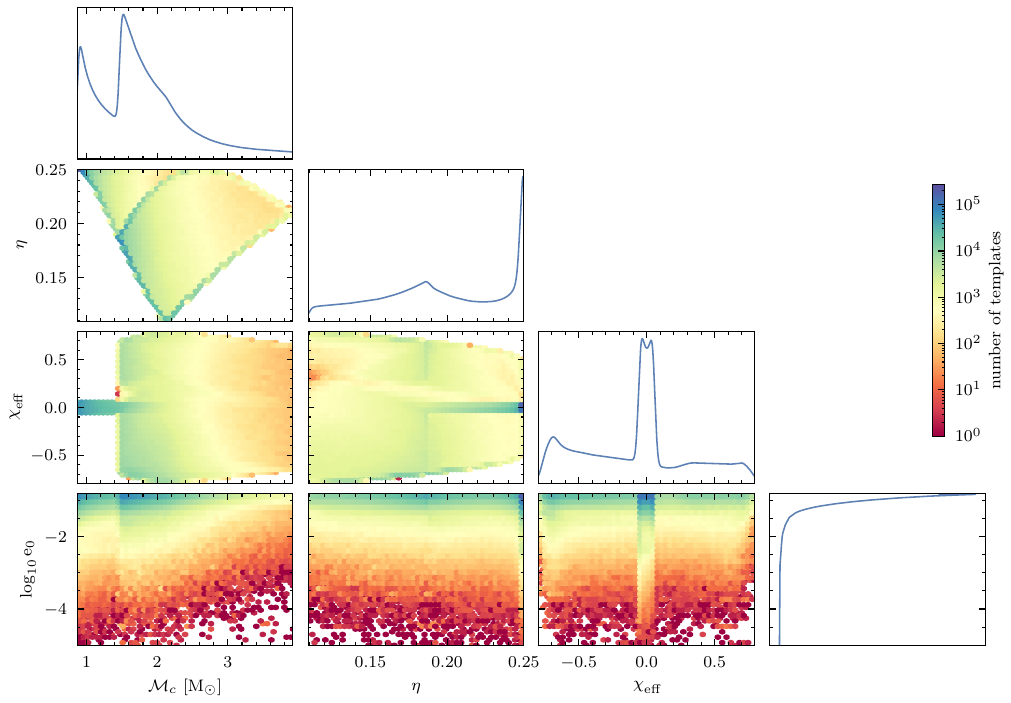}
\caption{\label{fig:ecc_bank_chirp_mass}  Visualisation of the eccentric template bank in the hexagonal binning of $(\mathcal{M}_c, \eta, \chi_{\rm eff}, e_0)$--parameter space. The blue lines show the 1D marginal distributions of the template density for each parameter. The colors represent the number of templates in each hexagon.}
\end{figure*}

We provide a comprehensive visualization of the constructed geometric eccentric template bank that spans the parameter space of BNS and NSBH binaries in Figs.~\ref{fig:ecc_bank_component_mass} and \ref{fig:ecc_bank_chirp_mass}. Covering the parameter space as detailed in Table~\ref{tab:params}, our bank contains $\eccbanksizeThreeD$ templates.
Figure~\ref{fig:ecc_bank_component_mass} illustrates the correlations between different pairs of template bank parameters, including primary and secondary masses $\left(m_1^{}, m_2^{}\right)$, the aligned-spin components $(s^{}_{1z}, s^{}_{2z})$, and the initial eccentricities $e_0^{}$. 
The blue lines denote the marginal distributions of the template density, providing insight into the distribution of templates across individual parameters. The hexagons are color-coded to indicate the density of templates in specific parameter regions, as quantified by the accompanying colorbar. Figure~\ref{fig:ecc_bank_chirp_mass} visualizes the template bank in a different parameter space defined by the chirp mass $\mathcal{M}_c= (m_1+m_2)\eta^{3/5}$, the symmetric mass ratio $\eta= m_1 m_2/(m_1+m_2)^2$, the effective inspiral spin $\chi_{\rm eff}$~\cite{Ajith:2009bn}, and the initial eccentricity $e_0$. 
Like in Fig.~\ref{fig:ecc_bank_component_mass}, we use blue marginal density lines and a color-coded scale to reflect the template distribution. 
We find that the density of templates increases sharply near the edges of the bank, illustrating the effects of the boundaries of the physical parameter space when mapping lattice points back from the $\xi$-coordinates to the physical parameters. Such an increase in template density is also seen at the transition points of BNS to NSBH systems. A refinement of the boundary effects could potentially reduce the number of templates, which we leave for future work.

%%%%%%%%%%%%%%%%%%%%%%%%%%%%%%%%%%%%%%%%%%%%%%
\section{Results}
\label{sec:performance}
%%%%%%%%%%%%%%%%%%%%%%%%%%%%%%%%%%%%%%%%%%%%%%
To assess the performance of the eccentric bank in detecting GW signals, we compute the \emph{effectualness}~\cite{Damour:1997ub} or \emph{fitting factor} ($\FF$)~\cite{Apostolatos:1995pj}, which is an indicator for the efficiency of a bank in detecting signals. We consider the mock signal population, referred to as \emph{injections}, to have moderate orbital eccentricities and spins (anti-)aligned with the orbital angular momentum. 
We also evaluate the effectualness of a quasi-circular bank constructed for the same mass--spin parameter space with the 3D $A_n^*$-lattice placement, using the geometric method developed in~\cite{Brown:2012qf, Harry:2013tca} against injections. We use the {\tt PyCBC} toolkit~\cite{alex_nitz_2024_10473621} to generate the quasi-circular bank.

In order to evaluate the effectualness of banks, we perform bank simulations, i.e., Monte Carlo studies where a large number of random, statistically independent injections $\{\hs\}$ are drawn, and for each injection the $\FF$ against the bank $\{\hb \}$ is determined. The $\FF$ is computed by maximizing the match of $\hs$ over the template bank $\{\hb\}$ and given as
\begin{equation}
\label{eq:ff}
\FF(\hs) = \max_{h \in \{\hb\}}  \mathrm{M}(\hs,h).
\end{equation}
The $\FF$ is sensitive to any disagreement between the waveform approximants~\cite{Nitz:2013mxa}. Therefore, to avoid systematics in our comparison we use the quasi-circular PN approximant {\tt TaylorF2} for quasi-circular bank and injections, and {\tt TaylorF2Ecc} for the eccentric bank and injections to disentangle the loss in effectualness due to discreteness and non-inclusion of eccentricity in the bank from the loss due to underlying differences in the waveforms approximants. We also keep the relevant PN orders, both in amplitude and phase, for injections and templates the same. 

For injections distributed uniformly in the sky, the event rate scales with the cube of the SNR, and thus, the fractional reduction in the detection of signals due to the discreteness of the template bank is proportional to  $1-\FF^3$.  A $\FF = 0.97$ thus corresponds to an approximate $10\%$ loss in the number of detections compared to an optimal search with  $\FF = 1$.

The SNR of an injection depends not only on the intrinsic parameters but also its extrinsic ones. Consequently, also the fitting factor depends on both sets of parameters, indicating vulnerability to selection effects~\cite{Ajith:2012mn}. 
In order to account for selection effects in bank simulations, we draw injections $\{ \hs \}$ covering both intrinsic and extrinsic parameters to compute the $\FF$ for each injection and then  evaluate the \emph{effective fitting factor} $\FFEF$~\cite{Buonanno:2002fy,Harry:2016ijz} defined as
\be
\label{eq:eff}
  \FFEF = \left( \frac{\sum_{i}\mathrm{FF}^3(h_s^i) \ \sigma^3(h_s^i)}{\sum_{i}\sigma^3(h_s^i)}\right)^{1/3},
\ee
where $\sigma(\hs)$ is proportional to the SNR of an optimally oriented (face-on) and located (overhead)  source at a fixed luminosity distance.
Hence, the effective fitting factor $\FFEF$ describes the average fraction of SNR recovered by the discrete template bank from an observed population of binary sources.

For the bank simulations, we draw $60,000$ random injections  covering component masses in the range $m^{}_1 \in \mathrm{U}(1,\,7) M_\odot$, $m^{}_2 \in \mathrm{U}(1,\,3)M_\odot $, BH spin magnitude $\in \mathrm{U}(0,\,0.9)$, NS spin magnitude $\in \mathrm{U}(0,\,0.05)$ and eccentricity $e^{}_0 \in \mathrm{U}(10^{-5}, 0.15)$ defined at 15 Hz.  
An additional $60,000$ quasi-circular injections are drawn from the same distributions but with zero eccentricity. We consider any binary component with mass $\leq 3\, M_\odot$ to be a neutron star. 
For both quasi-circular and eccentric injections, we consider the cosine of the angle $\iota$ describing the relative orientation of the total angular momentum of the binary with respect to the line-of-sight to be uniformly distributed in the interval $\mathrm{U}(0, 1)$, while the polarization angle $\psi$ is uniformly distributed in $\mathrm{U}(0, \pi)$. The sky location angles are drawn such that the injections are distributed uniformly in the sky, i.e., the right ascension $\phi \in \mathrm{U}(0, 2\pi)$ and the cosine of the declination $\cos\theta \in \mathrm{U}(0, 1)$.
A summary of the various bank simulation parameters is given in Table~\ref{tab:BankSimParams}. For our bank simulations and for generating the banks, we use the optimistic, publicly available LIGO O4 sensitivity noise curve {\tt aligoO4\_high.txt}~\cite{noise_curves} that has a horizon distance of 190 Mpc for canonical BNS with $m^{}_{1,2}=1.4 M_\odot$. 

\begin{table}[t]
\begin{center}
\begin{tabular}{c@{\quad}c}
\toprule
\toprule
Bank simulation parameter & Value \\
\midrule
Waveform approximant: &       \begin{tabular}{@{}c@{}}quasi-circular: {\tt TaylorF2}  \\   eccentric: {\tt TaylorF2Ecc}   \end{tabular}  \\
Primary mass $(m_1^{})$ & $\mathrm{U} (1,\,7)\,M_\odot$\\
Secondary mass $(m_2^{})$: & $\mathrm{U} (1,\,3)\,M_\odot$ \\
BH spin magnitude & $\mathrm{U}(0,\,0.9)$ \\
NS spin magnitude & $\mathrm{U}(0,\,0.05)$ \\
Eccentricity ($e_0$ at 15 Hz) & $\mathrm{U}(10^{-5},\,0.15)$ or 0 \\
Cosine of declination  $( \cos \theta)$ & $\mathrm{U} (-1, 1)$\\
Right ascension $(\phi)$ & $\mathrm{U}(0,\,2\pi)$ \\
Cosine of inclination angle $(\cos \iota)$ & $\mathrm{U} (0,\,1)$ \\
Polarization angle  $(\psi)$ & $\mathrm{U}  (0,\,2\pi)$ \\
Luminosity distance $(d_L)$ & 10 Mpc \\ 
Lower-frequency cutoff $(\flow)$ & $15\,$Hz \\
PSD & {\tt aligoO4\_high.txt}~\cite{noise_curves}\\
\bottomrule
\bottomrule
\end{tabular}
\end{center}
\caption{Parameters used for the bank simulations of eccentric and quasi-circular binaries.  The quasi-circular and eccentric binaries share the same parameter values in this table except for parameters relevant to the orbit type of binaries.    A binary component is a neutron star (black hole)  if its mass is $\leq (>) 3\, M_\odot $. In the table, $"\mathrm{U}"$ denotes uniform distribution.}
\label{tab:BankSimParams}
\end{table}%

We use two different template banks for our bank simulations: 
\begin{enumerate}
    \item An aligned-spin, quasi-circular template bank generated using a 3D $A_n^*$ lattice.
    \item An  aligned-spin eccentric template bank generated using a 3D    $A_n^*$ lattice. 
\end{enumerate}
The bank simulations are performed using  the workflow generator, {\tt  pycbc\_make\_bank\_verifier\_workflow} available in {\tt PyCBC}~\cite{alex_nitz_2024_10473621} with minor modifications in it. A selection of figures of merit for our banks is presented in Table~\ref{tab:banksimSummary}. 

\begin{table*}[bht]
\centering
\begin{ruledtabular}
\begin{tabular}{ccccccc}
\multirow{2}{*}{Bank Type }\ \ \    &	\multirow{2}{*}{Placement Lattice} \hspace{1mm} & \multirow{2}{*}{Bank Size} \hspace{1mm}  & \multirow{2}{*}{Injected Binary System }\ \ \  & \multicolumn{3}{c}{$\FF \leq 0.97$ [\%]}   \\
\cmidrule[0.8pt](lr{0.75em}){5-7}
& &  & &BNS  & NSBH & ALL  \\
\midrule[0.8pt]
   \multirow{2}{*}  {quasi-circular, AS  } &   \multirow{2}{*}{$A^*_3$}&   \multirow{2}{*}{\qcbanksizeThreeD}&quasi-circular, AS \ \ \ &0.01 &0.80 &0.67  \\ 
 &  & &eccentric, AS  \ \ \ &75.92 &68.95 &70.11 \\  \hline
    \multirow{2}{*}  {eccentric, AS  } &   \multirow{2}{*}{$A^*_3$}&   \multirow{2}{*}{\eccbanksizeThreeD}&quasi-circular, AS \ \ \ &1.28 &4.26 &3.76  \\ 
 &  & &eccentric, AS  \ \ \ &2.17 &6.38 &5.68 \\
\end{tabular}
\end{ruledtabular}
\caption{\label{tab:banksimSummary}Summary bank simulations with different types of aligned-spin (AS) injections on eccentric  and quasi-circular template banks. The first column specifies injections by their orbits, the second column indicates whether eccentricity is included in the bank's degrees of freedom. The third column describes the lattice placements used in the bank constructions, with  the notation  $A_3^*$-lattice referring to the 3D $A_n^*$-lattice. The fourth column lists the bank sizes, while the fifth column and onwards present the percentage of binaries with  $\FF$ values lower or equal to $0.97$. The label, `ALL` indicates the combined samples of BNS and NSBH injections for each orbit type.} 
\end{table*}

As discussed before, there will inevitably be a difference between the source's true parameters and those of the best-fit template due to the sparseness of the bank, leading to a loss in the number of detectable events. This occurs even if the source is within the search parameter space and the template waveform family accurately represents GW signals. 
Furthermore, inaccuracies in the template waveform compared to true GW signals, which arise from the exclusion of physical effects such as eccentricity -- our main focus here -- can further reduce the detection rate by degrading the filtered SNR or $\FF$. Therefore, we expect the effectualness of a quasi-circular template bank in detecting eccentric signals to be low.

\begin{figure*}[th!]
\centering
\includegraphics[width=0.48\textwidth]{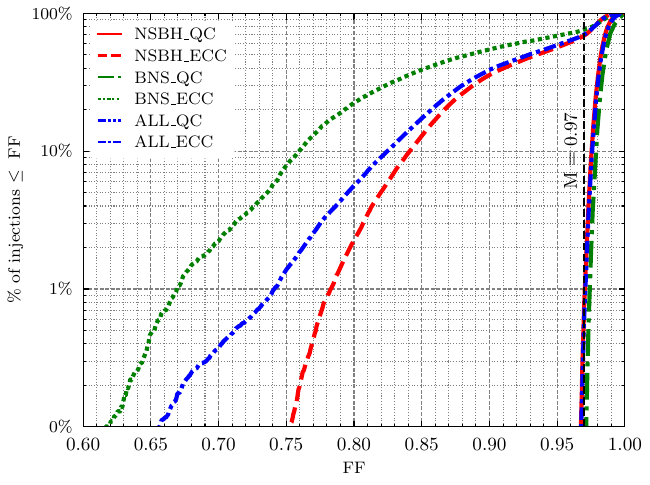}
\includegraphics[width=0.48\textwidth]{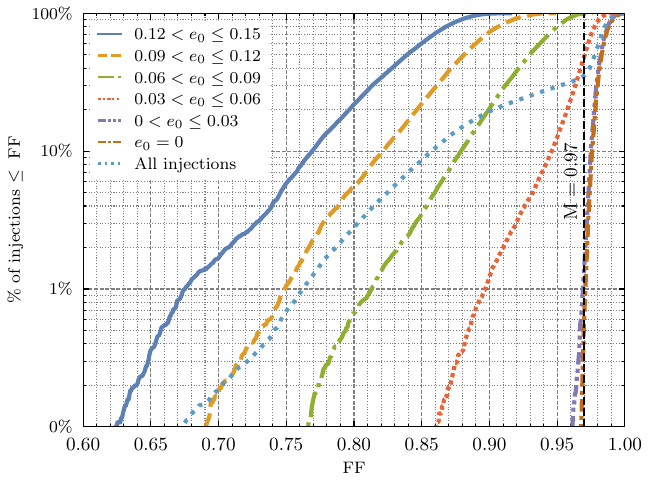}
\caption{\label{fig:qc_bank_ff_1} 
Performance of the quasi-circular bank. 
\emph{Left:} Each curve represents a cumulative histogram of FFs for a specific class of binaries, labelled `BNS`, `NSBH`, and `ALL` to indicate the binary neutron stars subpopulation, neutron star-black hole subpopulation, and a union of both subpopulations, respectively. The suffixes `QC` and `ECC` refer to the quasi-circular and eccentric injections, respectively. The vertical black line indicates the minimal match of the template bank, $\mathrm{MM}=0.97$. The bank is very effectual for quasi-circular binaries, while a significant loss in the bank's performance arises for eccentric binaries, with the worst fitting factors obtained for BNS sources. 
\emph{Right:} The $\FF$ binned into different ranges of eccentricity, showing how the efficiency of the quasi-circular template bank depends on eccentricity.}
\end{figure*}

In the left panel of Fig.~\ref{fig:qc_bank_ff_1} we show the fitting factor of the quasi-circular bank for eccentric and quasi-circular injections. We find that the quasi-circular bank is highly effectual in detecting the quasi-circular injections, i.e., almost all quasi-circular injections have $\FF \geq 0.97$. Conversely, for the eccentric injections we find that $\sim 70\%$ of injections have a $\FF \leq 0.97$ and are therefore missed by the bank. We see comparatively poorer performance of the quasi-circular bank for eccentric BNS sources than eccentric NSBH systems. We note that we allowed each injection to explore the entire parameter space of the bank before finding the best match template, rather than restricting the search to a narrow region, further emphasising the ineffectualness of the quasi-circular bank to detect signals from eccentric BNS and NSBH sources. 

To further highlight the poor performance of the quasi-circular bank to recover eccentric signals, we combine all eccentric and quasi-circular injections into one injection set and group them in bins of eccentricity.    
The $\FF$s corresponding to those bins are shown in the right panel of Fig.~\ref{fig:qc_bank_ff_1}.  Nearly one-third of the combined injections (black solid curve) have a fitting factor lower than the minimal match threshold of the bank, $\mathrm{MM}=0.97$. The bank performs similarly well for binaries with circular orbits and those with eccentricities in the range $0 < e_0 \leq 0.03$. However, it is significantly less effectual for binaries with eccentricities larger than $e_0>0.03$: For binaries with eccentricities in the range $0.03 < e_0 \leq 0.06$, 50\% of injections have a fitting factor below $0.97$.
For binaries with $e_0 > 0.06$, nearly all injections are missed by the quasi-circular bank, despite allowing each injection to explore the full parameter space before finding the best match template.   

Our results confirm that the quasi-circular template bank constructed with the {\tt TaylorF2} metric is highly ineffectual for detecting eccentric low-mass binaries. Ignoring eccentricity in the bank significantly decreases the detection rate of eccentric binaries, which imposes a strong selection bias against such systems, and only systems with with eccentricities below $0.06$ remain detectable.    

We now assess the effectualness of our eccentric bank (see Sec.~\ref{sec:bank}) by repeating the bank simulations using the same sets of quasi-circular and eccentric injections.
We note that our eccentric bank is $\sim 8.4$ times larger than the quasi-circular bank, while both banks were constructed with a minimal match of $\mathrm{MM}=0.97$.  

\begin{figure}[h]
\includegraphics[width=0.5\textwidth]{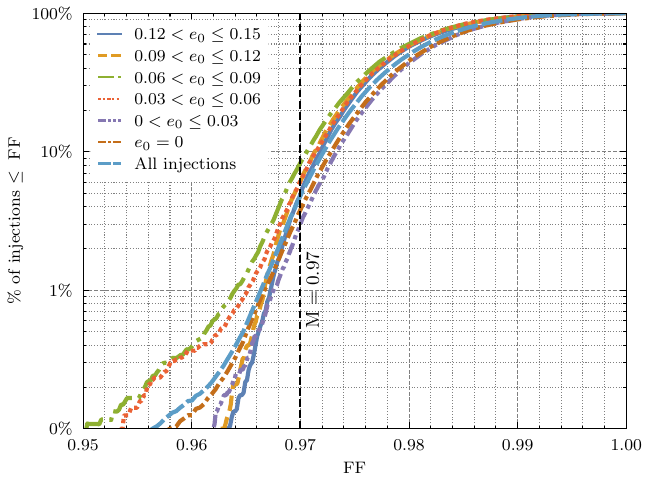}
\caption{\label{fig:fitting_factor_ecc_bank} Performance of the eccentric bank. We show the fitting factor for the same populations of injections as for the quasi-circular bank (solid black curve), and its dependence on the eccentricity grouped in bins.}
\end{figure}

Figure~\ref{fig:fitting_factor_ecc_bank} shows the fitting factor of the eccentric bank for all injections (solid black curve) and grouped into eccentricity bins. Firstly, we find that our eccentric bank is highly effectual in detecting both quasi-circular and eccentric injections, with $\lesssim 5 \%$ of injections having a fitting factor below $0.97$ and only litte dependence on the value of eccentricity. This is an order of magnitude improvement in comparison to the quasi-circular bank, where $\sim 35\%$ of the injections recorded a fitting below $0.97$. Secondly, we find that the worst fitting factor across all injections is still above $0.95$, while the fitting factor of the quasi-circular bank has a long tail that extends down to values as low as $\sim 0.62$, especially for eccentric BNS signals. Thirdly, while the eccentric bank is overall significantly more effectual, we see a slight degradation for the quasi-circular injections of which $\sim 4\%$ are missed by the eccentric bank. The largest number of missed injections are within the NSBH subpopulation. 

\begin{figure}[h]
\includegraphics[width=0.5\textwidth]{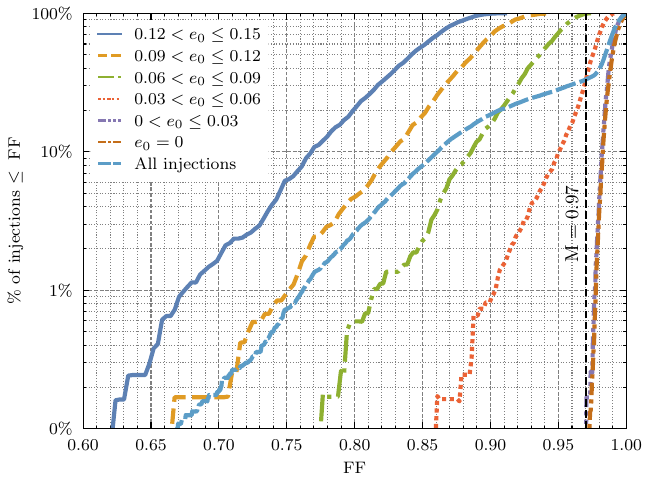}
\caption{\label{fig:fitting_factor_ecc_zero_bank} Fitting factor of the eccentric bank, where the eccentricities of the templates have been set to zero. Grouped by eccentricity, the $\FF$ distributions are similar to those of the quasi-circular bank (see right panel of Fig.~\ref{fig:qc_bank_ff_1}), demonstrating that the improved effectualness of the eccentric bank is not due to the higher number of templates. 
}
\end{figure}

As the eccentric bank is significantly larger than the quasi-circular bank, it is in principle possible that the improved effectualness is a byproduct of the increased number of templates rather than due to the inclusion of eccentricity. To determine whether this is the case, we take our eccentric bank but set $e^{}_0=0$ for all templates, effectively turning our eccentric bank into a densely-packed quasi-circular bank, and recompute the FF against the injections. The results are shown in Fig.~\ref{fig:fitting_factor_ecc_zero_bank}. Notably, comparing these results to the right panel of Fig.~\ref{fig:qc_bank_ff_1}, we obtain qualitatively similar results, suggesting strongly that the improvement in the recovery of eccentric signals is indeed due to the inclusion of eccentricity in the bank and not due to the larger bank size.

\begin{figure*}[h]
\centering
\includegraphics[width=0.48\textwidth]{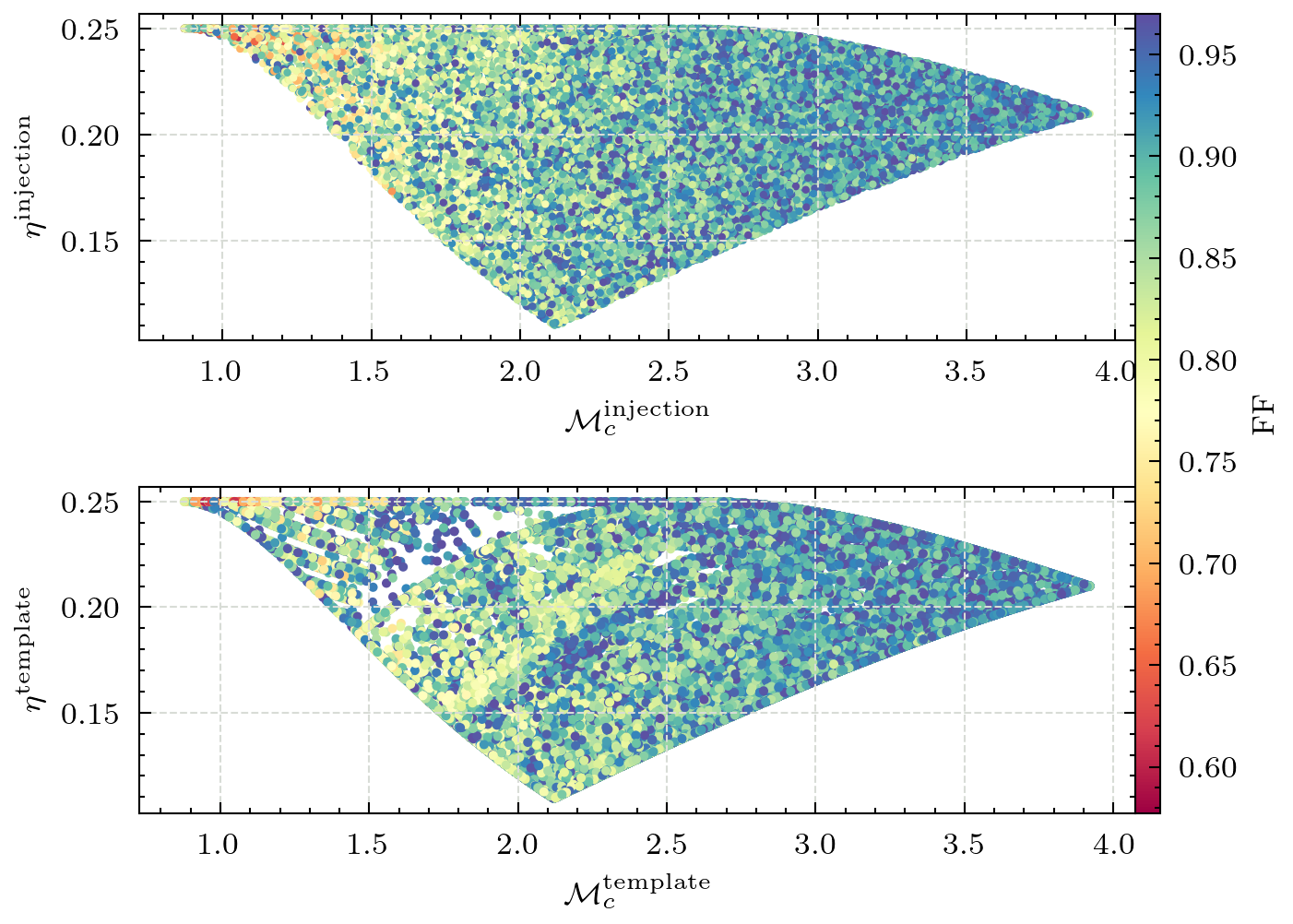}
\includegraphics[width=0.48\textwidth]{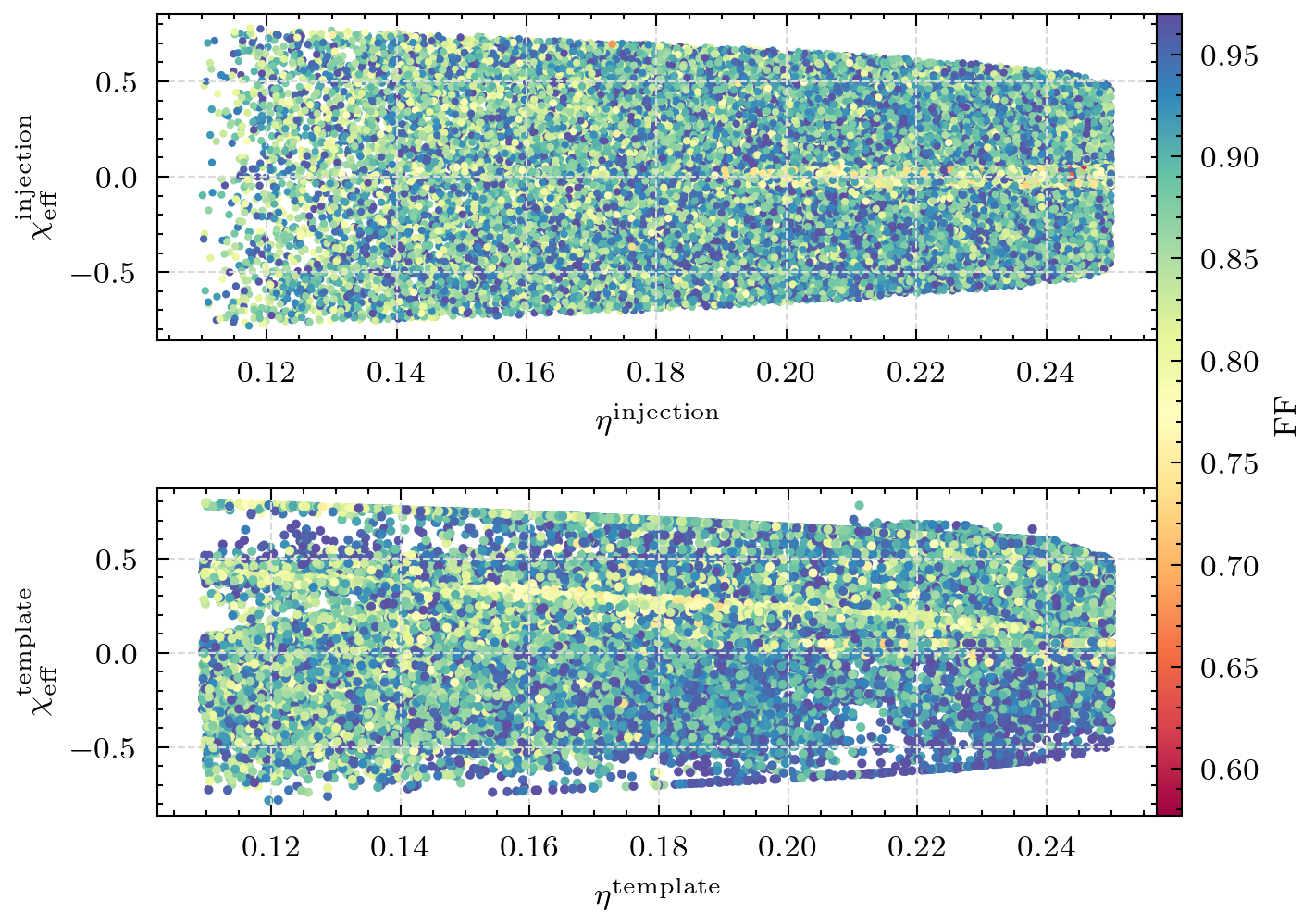}
\caption{\label{fig:qc_bank_samples_ff_lower_0p97} Impact of boundary effects. 
\emph{Top}: Mass and spin parameter of eccentric injections that are recovered with a  $\FF<0.97$ by the quasi-circular bank. 
\emph{Bottom}: Best-matching template parameters for the injections shown in the top panels. There is no significant difference in the $\FF$ distribution between injections close to the boundary or from the interior, implying that the neglect of eccentricity in the bank is the limiting factor.}
\end{figure*}

Furthermore, given the results presented in Figs.~\ref{fig:qc_bank_ff_1} and~\ref{fig:fitting_factor_ecc_zero_bank}, 
one might argue that injections with parameters near the edges of a template bank's parameter space constitute a significant fraction of $\FF < 0.97$ as they are less likely to match neighbouring templates, compared to injections far from the boundaries. 
However, this is not the case for our banks as is shown in Fig.~\ref{fig:qc_bank_samples_ff_lower_0p97}: The top panels show the chirp mass $\mathcal{M}_c$, symmetric mass ratio $\eta$ and effective spin $\chi^{}_{\rm eff}$ of all injections that are recovered with a $\FF < 0.97$ by the quasi-circular bank. The parameters of the corresponding best matching templates are shown in the bottom panels. 
It is clearly seen that the injections with poor $\FF$ and the best-matching templates are distributed relatively uniformly across the parameter space of the quasi-circular bank with no particular degradation towards the edges. The subtle stripes that can be seen arise from the internal boundary at the transition from NS to BH systems, where we find relatively low $\FF$s compared to other regions of the bank.
This indicates that boundary effects are subdominant and supports the conclusion that the poor recovery of these injections is due to the lack of an extra degree of freedom in the bank that captures the effect of eccentricity. 

\begin{figure*}[t]
\centering
\includegraphics[width=0.48\textwidth]{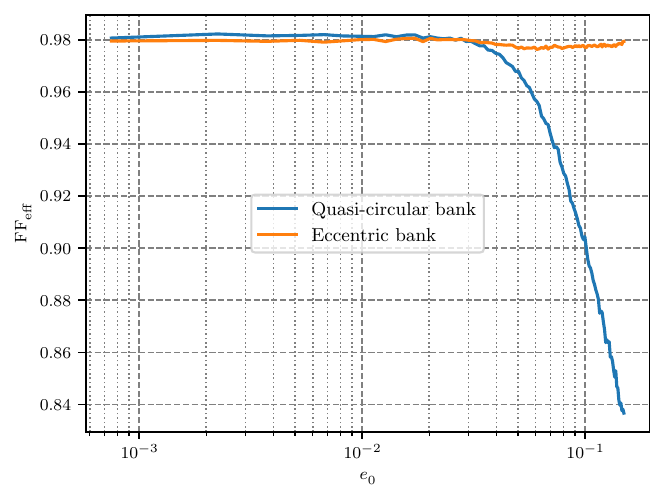 }
\includegraphics[width=0.48\textwidth]{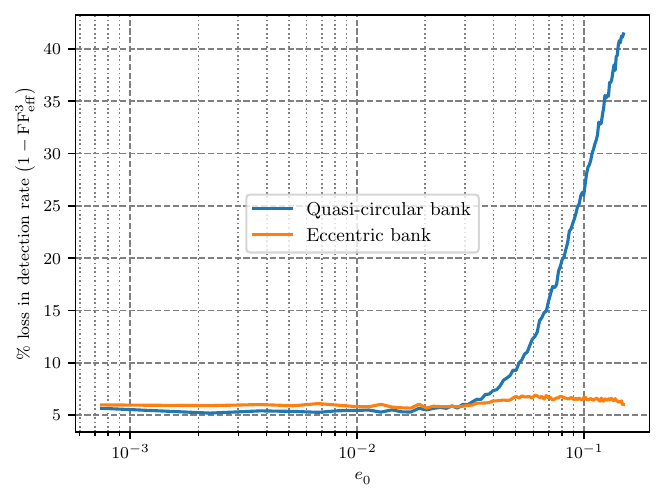 }
\caption{\label{fig:effective_ff_comparison} {\it Left}: Effective fitting factor $\FFEF$ of the eccentric (orange) and quasi-circular (blue) bank as a function of the eccentricity of the injections. The $\FFEF$ is computed over $100$ eccentricity bins with a bin size of $\Delta e^{}_0  = 0.0015$. {\it Right}: Loss in the detection rate as a function of eccentricity.}
\end{figure*}

Finally, as discussed around Eq.~\eqref{eq:eff}, to account for selection effects in our bank simulations and determine the loss in detection rate, we also compute the effective fitting factor $\FFEF$ for both banks as a function of the injected eccentricity. The results are shown in Fig.~\ref{fig:effective_ff_comparison}. We find comparable performance of the two banks for eccentricities $e_0 \leq 0.03$. For larger values of eccentricity, the effective fitting factor of the quasi-circular bank decays rapidly, while the one for the eccentric bank remains approximately constant. The right panel of the figure show the loss in detection rate. Again, we find comparable performance for $e_0\leq 0.03$ with a loss in the detection rate of $\lesssim 6\%$ for both banks. While this remains the case for the eccentric bank for the entire range of $e_0$ considered here, the quasi-circular bank incurs a loss in the detection rate of more than $25\%$ for $e_0 \leq 0.1$ that increases to more than $40\%$ for the highest values of $e_0$.

In summary, we find that a quasi-circular bank performs comparably to the eccentric bank for low-mass binaries with orbital eccentricities up to at most $0.03$ when entering the low-frequency end of current generation ground-based detectors. For higher values of eccentricity, the quasi-circular bank becomes highly ineffectual, failing to detect more than $40\%$ of signals, while the eccentric bank incurs a constant loss of $\lesssim 6\%$ due to the discreteness of the bank.

%%%%%%%%%%%%%%%%%%%%%%%%%%%%%%%%%%%%%%%
\section{Discussion and Conclusions}
\label{sec:discussion}
%%%%%%%%%%%%%%%%%%%%%%%%%%%%%%%%%%%%%%%
Observations of compact binaries on eccentric orbits offer a unique opportunity to probe different formation pathways and environments of astrophysical systems~\cite{Zevin:2021rtf}. The discovery of BNSs or NSBH binaries with measurable residual eccentricity would reveal novel insights into the binary properties at the time of formation~\cite{Fumagalli:2024gko}.

Current searches for modelled GW signals from compact binaries primarily use aligned-spin, quasi-circular template banks. 
While these searches are also successful in detecting binaries from an extended parameter space, higher-order modes~\cite{LIGOScientific:2020stg}, precession~\cite{Hannam:2021pit} and matter effects~\cite{LIGOScientific:2017vwq}, their efficacy is limited.
Therefore, search methods need to be improved to realize the full potential of GW astronomy in discovering diverse astrophysical sources.  

Developing an effectual and computationally efficient template bank that extends beyond aligned-spin, quasi-circular binaries has been a long-standing problem in GW data analysis, see e.g. Refs.~\cite{Capano:2013raa, Capano:2016dsf, McIsaac:2023ijd}. Recently, progress has been made on constructing the first template banks for eccentric binaries for low-mass eccentric binaries using the stochastic method~\cite{Nitz:2019spj,Dhurkunde:2023qoe, Lenon:2021zac}. 

In this paper, we constructed the first geometric template bank for spinning BNSs and NSBH binaries on moderately eccentric orbits by deriving an effective metric for the {\tt TaylorF2Ecc} inspiral approximant (Sec.~\ref{sec:metric}). We then demonstrated the severe limitations of a quasi-circular bank, spanning an equivalent mass and spin parameter space, in detecting eccentric signals with a loss of $\gtrsim 40\%$ of signals. 
In contrast, our eccentric bank misses $\lesssim 6\%$ of signals due the finite template spacing. Moreover, our bank is significantly smaller than a stochastic bank covering a similar parameter space.

In order to quantify the effects of neglecting eccentricity in GW searches,  we evaluated the FFs between GW signals from a fiducial population of moderately eccentric BNS and NSBH systems and an aligned-spin, quasi-circular template bank constructed with {\tt TaylorF2}. This study revealed that the quasi-circular template bank is efficient in detecting signals with eccentricities up to $e_0 \lesssim 0.03$ with an incurred loss of $6\%$ of signals, which is comparable to the losses due to bank discretization resulting from $\rm MM = 0.97$. 
However, for binaries with $e_0 > 0.03$, the quasi-circular bank's detection efficiency declines sharply, with up to $30\%$ of signals with eccentricities in the range $0.03 < e_0 \leq 0.06$ having fitting factors below the MM threshold. 
For binaries with $e_0 > 0.06$, up to $40\%$ of signals are lost, indicating a severe reduction in the detection capability of template bank-based searches when eccentricity is not accounted for, even for moderate eccentricities.

The geometric template bank constructed in Sec.~\ref{sec:bank} addresses the inefficiency of quasi-circular banks by incorporating eccentricity directly into the bank. 
Our bank construction approach is based on the method presented in Refs.~\cite{Brown:2012qf, Harry:2013tca} to construct a template bank for {\tt TaylorF2} combined with the PN parametrization of the {\tt TaylorF2Ecc} phase introduced in Sec.~\ref{sec:wf}, a global coordinate transformation and dimensional reduction to derive an effective metric for {\tt TaylorF2Ecc}. 
Using the efficient and flat metric derived in Sec.~\ref{sec:metric}, we optimally place templates in an extended parameter space that includes mass, (anti-)aligned spin, and eccentricity to construct a bank for BNS and NSBH systems. The geometric method ensures that our bank has comprehensive coverage of the targeted binary parameter space while minimizing the number of templates needed for an effective search. Our geometric bank for eccentric, aligned-spin binaries is less than one order of magnitude larger than the quasi-circular bank covering the same non-eccentric region of the parameter space. The efficiency offered by the geometric method addresses to some degree the computational challenges posed by searches involving eccentricity.

We note that our bank construction does not account for the angular parameters of  an eccentric orbit such as the argument of the periapsis or the mean anomaly, which could slightly alter the waveform morphology, however, the {\tt TaylorF2Ecc} waveform does not model these effects. The contributions from orbital orientation effects decay rapidly, and within the regime of {\tt TaylorF2Ecc}'s validity, omitting these angular parameters results in a maximum dephasing of approximately 0.2 radians~\cite{Moore:2016qxz}. Furthermore, this waveform also does not account for eccentricity-induced higher harmonics beyond the dominant harmonic ~\cite{Moreno-Garrido}, which can become comparable to the dominant harmonic in highly eccentric binaries~\cite{Peters:1963ux,Moore:2018kvz}. The lack of accurate closed-form waveform models incorporating the orientation of the orbit and higher-order harmonics limits the exploration of the geometric eccentric template bank construction method in the high-eccentricity regime as well. The challenges in going beyond the dominant harmonic are expected to be similar to those encountered when attempting to construct template banks with higher-order modes for quasi-circular binaries~\cite{Capano:2013raa} or precession~\cite{Privitera:2013xza, McIsaac:2022odb}.
Future work will focus on addressing these limitations by by using more complete waveforms, which could refine and extend the detection efficiency of searches to a wider parameter space of astrophysical signals. Furthermore, highly asymmetric binaries, especially in the NSBH region, will likely have strong higher-order modes, which are not incorporated in the bank presented here, leaving room for future improvement. 

A significant computational bottleneck in the template generation method presented here and in~\cite{Brown:2012qf,Harry:2013tca} lies in the inverse mapping from lattice points in the principal component space to the physical parameter space, as no analytical solution exists. This is typically resolved using brute-force methods that involve numerous trials to locate corresponding physical points, making the process computationally expensive. While we employ a binary tree approach to reduce the number of trials, the bottleneck could be further addressed with advanced techniques such as machine learning algorithms like UMAP~\cite{2018arXiv180203426M}) or autoencoders~\cite{autoencoder}, which may offer more efficient mappings. We leave this as an avenue for future exploration.

Finally, incorporating the template bank into large-scale searches on detector data will be crucial for fully assessing its impact on the detection of eccentric binaries. We leave this for future work. Beyond its applications in searches, the metric introduced here has the potential to facilitate the development of rapid sampling algorithm for reliable parameter estimation similarly to Ref.~\cite{Nitz:2024nhj}, possibly having a significant impact on low-latency GW observations.

%%%%%%%%%%%%%%%%%%%%%%%%%%%%%%%%
\section*{Acknowledgments}
%%%%%%%%%%%%%%%%%%%%%%%%%%%%%%%%
The authors thank Ian Harry for carefully reading the manuscript and providing helpful comments and suggestions. The authors also thank  Bhooshan Gadre, Soumen Roy and Tito Dal Canton for useful discussions. The authors are particularly thankful to Duncan Macleod for help with carrying out computations on the Hawk Computing Cluster.
K.S.P., G.P. and P.S. acknowledge support from STFC grant ST/V005677/1; G.P and P.S. also acknowledge support from STFC grant ST/Y00423X/1.
G.P. is very grateful for support from a Royal Society University Research Fellowship URF{\textbackslash}R1{\textbackslash}221500 and RF{\textbackslash}ERE{\textbackslash}221015. GP gratefully acknowledges support from an NVIDIA Academic Hardware Grant. P.S. also acknowledges support from a Royal Society Research Grant RG{\textbackslash}R1{\textbackslash}241327. 
Computations were performed using the University of Birmingham's BlueBEAR HPC service, which provides a High Performance Computing service to the University's research community, as well as on resources provided by Supercomputing Wales, funded by STFC grants ST/I006285/1 and ST/V001167/1 supporting the UK Involvement in the Operation of Advanced LIGO. 
This manuscript has the LIGO document number P2400566.

\onecolumngrid
\appendix
\section{PN  coefficients: Quasi-circular phase}
\label{app:pn_coeff}

The full expression of Eq.~\eqref{Eq:qc_phase1}  with non-zero terms is as follows

%\begin{widetext}
\be
\begin{split}
\Psi^{}_{\mathrm {QC}} (f; \vec{\rm \theta}_{\rm int}) =&~  \varphi^{}_0 (\vec{\rm \theta}_{\rm int})  f^{-5/3}  +  \varphi^{}_2 (\vec{\rm \theta}_{\rm int})  f^{-1}  + \varphi^{}_3 (\vec{\rm \theta}_{\rm int})  f^{-2/3} + \varphi^{}_4 (\vec{\rm \theta}_{\rm int})  f^{-1/3} + \varphi^{}_5 (\vec{\rm \theta}_{\rm int})  + \varphi^\ell_5 (\vec{\rm \theta}_{\rm int}) \log{f}   \\
&+ \varphi^{}_6(\vec{\rm \theta}_{\rm int})   f^{1/3}  + \varphi^\ell_6(\vec{\rm \theta}_{\rm int})  \log{f}  f^{1/3} + \varphi^{}_7(\vec{\rm \theta}_{\rm int})   f^{2/3} .
\end{split}
\ee
%\end{widetext}
% 
The frequency domain quasi-circular phase' is expanded to 3.5 PN order, including the quadratic-in-spin terms up to the 3PN order and the cubic-in-spin terms at  3.5PN order.  The PN coefficients $\{ \varphi^{}_i,   \varphi^{\ell}_i   \}$ are  given as ~\cite{Buonanno:2009zt,Arun:2008kb,Mishra:2016whh, Damour:2000zb,Damour:2002kr}

\begin{equation}
\varphi^{}_{0}  =  \frac{3}{128 (\mathcal{M} \pi )^{5/3}},
\end{equation}
\begin{equation}
\varphi^{}_{2}\, =\, \varphi^{}_{0} \,    \frac{ ( \mathcal{M} \pi) ^{2/3}}{\eta ^{2/5}}  \left[\frac{3715}{756}+\frac{55 \eta }{9}\right],
\end{equation}
\begin{equation}
\begin{split}
\varphi^{}_{3}\, = \, \varphi^{}_{0}\,     \frac{  \mathcal{M} \pi }{\eta ^{3/5}}  &\left[-16 \pi +\frac{113 \chi^{}_a \delta }{3} + \chi^{}_s \left(\frac{113}{3}-\frac{76 \eta }{3}\right)\right],
\end{split}
\end{equation}
\be
\begin{split}%%
\varphi^{}_{4}  = \, \varphi^{}_{0}     \frac{ ( \mathcal{M} \pi) ^{4/3}}{\eta ^{4/5}}  &\left[  \frac{15293365}{508032} + \frac{27145 \eta }{504} + \frac{3085\eta ^2}{72} -   \frac{405 \chi^{}_a \chi^{}_s \delta }{4} + \chi_s^2 \left(-\frac{405}{8}+\frac{5 \eta }{2}\right) + \chi_a^2 \left(-\frac{405}{8}+200 \eta \right)\right] ,
\end{split}
\ee
\be
\begin{split}%%
\varphi^{}_5   = \varphi^{}_{0}     \frac{ ( \mathcal{M} \pi) ^{5/3}}{\eta}  & \left[  \frac{38645 \pi }{756} -\frac{65 \pi  \eta }{9}     -\frac{140 \chi^{}_a \delta  \eta }{9}- \frac{732985 \chi^{}_a \delta }{2268}+\frac{340 \chi^{}_s \eta ^2}{9}  +\frac{24260 \chi^{}_s \eta }{81} -\frac{732985 \chi^{}_s}{2268}\right]\\
&   \times  \left[1+ \log{\left\{\frac{\mathcal{M}\pi}{\eta^{3/5}}\right\}}\right],
\end{split}
\ee%
\be
\begin{split}%
\varphi^\ell_5   =\varphi^{}_{0}     \frac{ ( \mathcal{M} \pi) ^{5/3}}{\eta}  & \left[  \frac{38645 \pi }{756} -\frac{65 \pi  \eta }{9}   - \frac{140 \chi^{}_a \delta  \eta }{9}- \frac{732985 \chi^{}_a \delta }{2268}+\frac{340 \chi^{}_s \eta ^2}{9}  +\frac{24260 \chi^{}_s \eta }{81}- \frac{732985 \chi^{}_s}{2268}\right],%
\end{split}
\ee
\be
\begin{split}%
\varphi^{}_{6}  =   \varphi^{}_{0}  \frac{ ( \mathcal{M} \pi) ^{2}}{\eta^{6/5}} & \left[  \frac{11583231236531}{4694215680} - \frac{640 \pi ^2}{3}  -  \frac{15737765635 \eta }{3048192}    +  \frac{2255 \pi^2 \eta }{12} +\frac{76055 \eta ^2}{1728}- \frac{127825 \eta ^3}{1296} -520 \pi  \chi^{}_s \eta\, + \right. \\
  &90 \pi  \chi^{}_s  +\chi^{}_a \left( \frac{2270 \pi \delta }{3}+\frac{75515 \chi^{}_s \delta  -\frac{8225 \chi^{}_s \delta  \eta }{18} }{144} \right)+ \chi_a^2 \left(  \frac{75515}{288} -  \frac{263245 \eta }{252} - 480 \eta ^2 \right) +  \\
& \left.    \chi_s^2 \left(\frac{75515}{288} -\frac{232415 \eta }{504} + \frac{1255 \eta ^2}{9}\right)   -\frac{6848 \gamma^{}_{\mathrm{E}} }{21}-\frac{13696 \log (2)}{21}-  \frac{6848}{63}  \log{\left\{ \frac{\mathcal{M}\pi}{\eta^{3/5}} \right\} }  \right],
\end{split}
\ee
\be%
\varphi^\ell_{6}  =- \varphi^{}_{0}     \frac{ ( \mathcal{M} \pi) ^{2}}{\eta ^{6/5}} \frac{6848}{63} ,
\ee

\be
\begin{split}%
\varphi^{}_{7}  =  \varphi^{}_{0}   \frac{ ( \mathcal{M} \pi) ^{7/3}}{\eta ^{7/5}}  & \left[  \frac{77096675 \pi }{254016} +\frac{378515 \pi  \eta }{1512}  - \frac{74045 \pi  \eta ^2}{756}  - \frac{25150083775 \chi^{}_s}{3048192} +\frac{10566655595 \chi^{}_s \eta }{762048} - \frac{1042165 \chi^{}_s \eta ^2}{3024} +\right. \\
&  \frac{5345 \chi^{}_s \eta ^3}{36} +   \chi^{}_a \left(  -\frac{25150083775 \delta }{3048192} +\frac{26804935 \delta  \eta }{6048}   -    \frac{215}{2} \chi_s^2 \delta  \eta +\frac{14585 \chi_s^2 \delta }{8}-\frac{1985 \delta  \eta ^2}{48}  \right)+\\
 & \left.  \chi_a^2 \left( \frac{14585 \chi^{}_s}{8}-7270 \chi^{}_s \eta + 80 \chi^{}_s \eta ^2\right)+ \chi_a^3 \left(\frac{14585 \delta }{24}-2380 \delta  \eta \right)+  \chi_s^3 \left(\frac{4585}{24}-\frac{475 \eta }{6}+\frac{100
   \eta ^2}{3} \right)\right],
\end{split}
\ee
where $\delta = \sqrt{1-4\eta}$, $\chi^{}_s=(\chi^{}_1+\chi^{}_2)/2$, $\chi^{}_a=(\chi^{}_1-\chi^{}_2)/2$. $\gamma^{}_E$ is the Euler constant, $\gamma^{}_E=0.5772156\cdots$.

\section{PN  coefficients: Eccentric phase}
\label{app:pn_coeff_ecc}

The full expression of Eq.~\ref{Eq:ecc_phase1}  is as follows
\be
\begin{split}
\Psi^{}_{\mathrm {Ecc}} (f; \vec{\rm \theta})  =& ~ \varepsilon^{}_{00} (\vec{\rm \theta}_{\rm int}) f_{\rm ecc}^{19/9}f^{-34/9} +   \varepsilon^{}_{20} (\vec{\rm \theta}_{\rm int}) f_{\rm ecc}^{19/9}f^{-28/9} +   \varepsilon^{}_{02} (\vec{\rm \theta}_{\rm int})  f_{\rm ecc}^{25/9}f^{-34/9} + \varepsilon^{}_{30} (\vec{\rm \theta}_{\rm int}) f_{\rm ecc}^{19/9}f^{-25/9}  + \\
& \varepsilon^{}_{03} (\vec{\rm \theta}_{\rm int})  f_{\rm ecc}^{28/9}f^{-34/9}  + \varepsilon^{}_{40}  (\vec{\rm \theta}_{\rm int}) f_{\rm ecc}^{19/9}f^{-22/9} + \varepsilon^{}_{22} (\vec{\rm \theta}_{\rm int}) f_{\rm ecc}^{25/9}f^{-28/9}  +  \varepsilon^{}_{04}  (\vec{\rm \theta}_{\rm int}) f_{\rm ecc}^{31/9}f^{-34/9}+ \\  
& \varepsilon^{}_{50} (\vec{\rm \theta}_{\rm int}) f_{\rm ecc}^{19/9}f^{-19/9} +   \varepsilon^{}_{23}  (\vec{\rm \theta}_{\rm int})f_{\rm ecc}^{28/9}f^{-28/9} +  \varepsilon^{}_{32} (\vec{\rm \theta}_{\rm int}) f_{\rm ecc}^{25/9}f^{-25/9} + \varepsilon^{}_{05} (\vec{\rm \theta}_{\rm int}) f_{\rm ecc}^{34/9}f^{-34/9}  + \\
&   \varepsilon^{}_{60} (\vec{\rm \theta}_{\rm int})  f_{\rm ecc}^{19/9}f^{-16/9} +    \varepsilon^{}_{24} (\vec{\rm \theta}_{\rm int})  f_{\rm ecc}^{31/9}f^{-28/9} +\varepsilon^{}_{33} (\vec{\rm \theta}_{\rm int})   f_{\rm ecc}^{28/9}f^{-25/9}     +  \varepsilon^{}_{42} (\vec{\rm \theta}_{\rm int})  f_{\rm ecc}^{25/9}f^{-22/9}  + \\
& \varepsilon^{}_{06} (\vec{\rm \theta}_{\rm int})  f_{\rm ecc}^{37/9}f^{-34/9} +  \varepsilon_{60}^\ell (\vec{\rm \theta}_{\rm int})  f_{\rm ecc}^{19/9}f^{-16/9}  +  \varepsilon_{06}^\ell (\vec{\rm \theta}_{\rm int})  \log{f}   f_{\rm ecc}^{37/9}f^{-34/9} .
\end{split}
\ee

As mentioned in the main text,  $\Psi^{}_{\mathrm {Ecc}}$  has no eccentricity-spin cross terms and coefficients of $f$ are at leading order in eccentricity $\mathcal{O}(e^2)$.  The eccentric PN coefficients~\cite{Moore:2016qxz} are
\be
  \varepsilon^{}_{0} =  -\frac{2355 e_0^2}{1462} \varphi^{}_0,
\ee
\be
   \varepsilon^{}_{20} =    \varepsilon^{}_{0}  \frac{ \left( \mathcal{M} \pi \right) ^{2/3} }{\eta ^{2/5}}  \left[\frac{299076223}{81976608}+\frac{18766963 \eta }{2927736}\right],
\ee
\be 
   \varepsilon^{}_{02}=   \varepsilon^{}_{0}  \frac{ \left( \mathcal{M} \pi \right) ^{2/3} }{\eta ^{2/5}}  \left[\frac{2833}{1008}-\frac{197 \eta }{36}\right],
\ee
\be
   \varepsilon^{}_{30}=  -   \varepsilon^{}_{0}  \frac{ \left( \mathcal{M} \pi^2 \right)  }{\eta ^{3/5}}   \frac{2819123 }{282600},
\ee
\be
   \varepsilon^{}_{03}=     \varepsilon^{}_{0}  \frac{ \left( \mathcal{M} \pi^2 \right)  }{\eta ^{3/5}}   \frac{377 }{72},
\ee
\be
\begin{split}
   \varepsilon^{}_{40} =    \varepsilon^{}_{0}  \frac{ \left( \mathcal{M} \pi \right) ^{4/3} }{\eta ^{4/5}} & \left[\frac{16237683263}{3330429696}+\frac{24133060753 \eta }{971375328} +  \frac{1562608261 \eta ^2}{69383952}\right],
 \end{split}
 \ee
 \be
 \begin{split}
    \varepsilon^{}_{22} =    \varepsilon^{}_{0}  \frac{ \left( \mathcal{M} \pi \right) ^{4/3} }{\eta ^{4/5}}  &\left[\frac{847282939759}{82632420864} -\frac{718901219 \eta }{368894736}  -  \frac{3697091711 \eta ^2}{105398496}\right],
 \end{split}
 \ee
  \be
 \begin{split}
     \varepsilon^{}_{04} =   \varepsilon^{}_{0}  \frac{ \left(\mathcal{M} \pi \right) ^{4/3} }{\eta ^{4/5}}  &\left[-\frac{1193251}{3048192} -\frac{66317 \eta }{9072}     +  \frac{18155 \eta ^2}{1296}\right],
  \end{split}
 \ee

 \be
 %\begin{split}
   \varepsilon^{}_{50}   =    \varepsilon^{}_{0}  \frac{ \left(\mathcal{M} \pi \right) ^{5/3} } {\eta }   \left[-\frac{2831492681 \pi }{118395270}-\frac{11552066831 \pi  \eta }{270617760}\right],
% \end{split}
 \ee
 
 \be
    \varepsilon^{}_{23} =   \varepsilon^{}_{0}  \frac{ \left(\mathcal{M} \pi \right) ^{5/3} } {\eta }  \left[\frac{112751736071 \pi }{5902315776}+\frac{7075145051 \pi  \eta }{210796992}\right],
  \ee
  
  \be
    \varepsilon^{}_{32} =   \varepsilon^{}_{0}  \frac{ \left(\mathcal{M} \pi \right) ^{5/3} } {\eta }   \left[-\frac{7986575459 \pi }{284860800}+\frac{555367231 \pi  \eta }{10173600}\right],
  \ee
  
  \be
      \varepsilon^{}_{05} =    \varepsilon^{}_{0}  \frac{ \left(\mathcal{M} \pi \right) ^{5/3} } {\eta }  \left[\frac{764881 \pi }{90720}-\frac{949457 \pi  \eta }{22680}\right],
  \ee
  \be
  \begin{split}
        \varepsilon^{}_{60} =   \varepsilon^{}_{0}  \frac{ \left(\mathcal{M} \pi \right) ^{2} } {\eta^{6/5} }  & \left[-\frac{43603153867072577087}{132658535116800000}+ \frac{536803271 \gamma^{}_{E}}{19782000} +  \frac{15722503703 \pi ^2}{325555200} +   \eta \left(   \frac{299172861614477}{689135247360}-   \right. \right.\\
      &  \left. -\frac{15075413 \pi ^2}{1446912}\right)  + \frac{3455209264991 \eta ^2}{41019955200}+\frac{50612671711
\eta ^3}{878999040}+   \frac{7064324789 \log 2}{59346000} - \frac{1121397129 \log 3}{17584000}+ \\
& \left. \frac{536803271}{19782000} \log \left\{  \frac{ ( \mathcal{M} \pi)^{1/3}} {\eta^{1/5}} \right\} \right],
  \end{split}
  \ee
  
  \be
    \varepsilon^\ell_{60} =   \varepsilon^{}_{0}  \frac{ \left(\mathcal{M} \pi \right) ^{2} } {\eta^{6/5} }  \frac{536803271}{59346000},
  \ee
  
  \be
  \begin{split}
     \varepsilon^{}_{24} =    \varepsilon^{}_{0}  \frac{ \left(\mathcal{M} \pi \right) ^{2} } {\eta^{6/5} }   &\left[-\frac{356873002170973}{249880440692736}-\frac{260399751935005 \eta }{8924301453312}  +\frac{150484695827
\eta ^2}{35413894656}+\frac{340714213265 \eta ^3}{3794345856}\right],
\end{split}
  \ee
  
  \be
      \varepsilon^{}_{33} = -   \varepsilon^{}_{0}  \frac{ \mathcal{M}^2 \pi^4  } {\eta^{6/5} }  \frac{1062809371}{20347200},
  \ee
  
  \be
  \begin{split}
    \varepsilon^{}_{42} =   \varepsilon^{}_{0}  \frac{ \left(\mathcal{M} \pi \right) ^{2} } {\eta^{6/5} }  & \left[   \frac{46001356684079}{3357073133568}+\frac{253471410141755 \eta }{5874877983744 }  -\frac{1693852244423 \eta
^2}{23313007872}-\frac{307833827417 \eta ^3}{2497822272}    \right],
  \end{split}
  \ee
  
  \be
  \begin{split}
      \varepsilon^{}_{06} =   \varepsilon^{}_{0}  \frac{ \left(\mathcal{M} \pi \right) ^{2} } {\eta^{6/5} }  & \left[  \frac{26531900578691}{168991764480}-\frac{3317 \gamma^{}_{E}}{126}  + \frac{122833 \pi ^2}{10368} + \eta \left( \frac{9155185261}{548674560}- \frac{3977 \pi ^2}{1152}\right)  - \frac{5732473 \eta ^2}{1306368}-\right. \\
&\left.  \frac{3090307 \eta ^3}{139968} -\frac{12091 \log 2}{1890}-\frac{26001 \log 3}{560}- \frac{3317}{126} \log \left\{ \frac{ \left(\mathcal{M} \pi \right)^{1/3} } {\eta ^{1/5}}  \right\}  \right],  
  \end{split}
  \ee
  
  \be
      \varepsilon^\ell_{06} =  -  \varepsilon^{}_{0}  \frac{ \left(\mathcal{M} \pi \right) ^{2} } {\eta^{6/5} } \frac{3317}{378}.
  \ee
%  \end{widetext}
\section{Phase with dimensionless coefficients}
\label{app:dimensionless_phase}
The phase $\Psi^{}_{\mathrm{F}2e} $ in terms of the dimensionless coefficients $\{\zeta^{}_i , \zeta_i^\ell,  \kappa^{}_i ,   \kappa^\ell_{6} \}$ as a power series in $x$:

\be
  \begin{split}
\Psi^{}_{\mathrm{F}2e}(x: \zeta^{}_i , \zeta_i^\ell,  \kappa^{}_i ,   \kappa^\ell_{6} ) &=  \Psi_{\mathrm {QC}} ^\prime(x;  \zeta_i , \zeta_i^\ell) + \Psi'_{\mathrm {Ecc}} (x: \kappa_i ,   \kappa^\ell_{6} )\\
& = -2 \phi^{}_0 + \zeta^{}_0  x^{-5/3} +  \zeta^{}_2  x^{-1} +  \zeta^{}_3  x^{-2/3}  + \zeta^{}_4  x^{-1/3}   +  \zeta^{}_6x^{1/3} + \zeta^{}_7  x^{2/3} +   \zeta^{}_8  x   +\zeta^{\ell}_5 \log x  + \\
&\quad      \zeta^{\ell}_6  \log x\, x^{1/3}  +   \kappa^{}_{0}  x^{-34/9} +   \kappa^{}_{2}  x^{-28/9}  +  \kappa^{}_{3}  x^{-25/9} +   \kappa^{}_{4}  x^{-22/9}+ \kappa^{}_{5}  x^{-19/9} + \kappa^{}_{6}  x^{-16/9}   +\\
&\quad  \kappa^\ell_{6} \log{x}\, x^{-16/9} .
\end{split}
\ee

The dimensionless coefficients $  \kappa^{}_{i},   \kappa^\ell_{i}$ are given as
\be
\begin{split}
 \kappa^{}_0 =&  \left(  \varepsilon^{}_{00}   f_{\rm ecc}^{19/9}  + \varepsilon^{}_{02}   f_{\rm ecc}^{25/9}  + \varepsilon^{}_{03}   f_{\rm ecc}^{28/9}    + \varepsilon^{}_{04}   f_{\rm ecc}^{31/9}   \varepsilon^{}_{05}   f_{\rm ecc}^{34/9}  + \varepsilon^{}_{06}   f_{\rm ecc}^{37/9} + \varepsilon_{06}^\ell \log{f^{}_{\rm ecc}}  f_{\rm ecc}^{37/9} \right) f_0^{-34/9} 
 \end{split}
\ee
\be
\begin{split}
 \kappa^{}_2 =&  \left(  \varepsilon^{}_{20}   f_{\rm ecc}^{19/9}  + \varepsilon^{}_{22}   f_{\rm ecc}^{25/9}  + \varepsilon^{}_{23}   f_{\rm ecc}^{28/9}    + \varepsilon^{}_{24}   f_{\rm ecc}^{31/9} \right)   f_0^{-28/9} 
 \end{split}
\ee
\be
\begin{split}
 \kappa^{}_3 =&  \left(  \varepsilon^{}_{30}   f_{\rm ecc}^{19/9}  + \varepsilon^{}_{32}   f_{\rm ecc}^{25/9}  + \varepsilon^{}_{33}   f_{\rm ecc}^{28/9}    + \varepsilon^{}_{24}   f_{\rm ecc}^{31/9} \right)  f_0^{-25/9} 
 \end{split}
\ee
\be
\begin{split}
 \kappa^{}_4 = \left(  \varepsilon^{}_{40}   f_{\rm ecc}^{19/9}  + \varepsilon^{}_{42}   f_{\rm ecc}^{25/9}  \right) f_0^{-22/9} 
 \end{split}
\ee
\be
\begin{split}
 \kappa^{}_5 =  \varepsilon^{}_{50}   f_{\rm ecc}^{19/9}   f_0^{-19/9} 
 \end{split}
\ee
\be
\begin{split}
 \kappa^{}_6 =  \left( \varepsilon^{}_{60}   +  \varepsilon_{60}^\ell   \log f_0  \right)   f_{\rm ecc}^{19/9}   f_0^{-16/9} 
 \end{split}
\ee
\be
\begin{split}
 \kappa_6^\ell =   \varepsilon_{60}^\ell    f_{\rm ecc}^{19/9}  f_0^{-16/9} 
 \end{split}
\ee
\twocolumngrid
%\end{widetext}
\bibliography{EccVsPrec}
\end{document}